\documentclass[aps,preprint,amsmath,amssymb,floatfix,longbibliography]{revtex4-2}
\usepackage{graphicx}  
\usepackage{dcolumn}  
\usepackage{bm}          
\usepackage[usenames,dvipsnames]{color}
\usepackage{ulem}
\DeclareMathOperator{\sign}{sign}

\begin{document}
\title{Active nematic flows confined in a two dimensional channel with hybrid alignment at the walls: a unified picture}
\author{C. Rorai$^{1}$, F. Toschi$^{2}$ and I. Pagonabarraga$^{1,3,4}$}
\affiliation{
$^1$CECAM, Centre Europ\'een de Calcul Atomique et Mol\'eculaire, \'Ecole Polytechnique F\'ed\'erale de Lausanne (EPFL), Batochime, Avenue Forel 2, 1015 Lausanne, Switzerland; $^2$Department of Applied Physics, Fluid Dynamics and Heat Transfer, Technical University of Eindhoven (TU/e); $^3$Departament de F\'{\i}sica de la Mat\`eria Condensada, Universitat de Barcelona, C. Mart\'{\i} Franqu\`es 1, 08028 Barcelona, Spain; $^4$University of Barcelona Institute of Complex Systems (UBICS), Universitat de Barcelona, 08028 Barcelona, Spain.}

\date{\today}
\begin{abstract}
Active nematic fluids confined in narrow channels are known to generate spontaneous flows when the activity is sufficiently intense. 
Recently, it was demonstrated [R. Green, J. Toner and V. Vitelli, {\it Phys. Rev. Fluids}, {\bf 2}:104201 (2017)] that if the molecular anchoring at the channel walls is conflicting: perpendicular on one plate and parallel on the other, flows are initiated  
even 
in the zero activity limit. An analytical laminar velocity profile for this specific configuration was derived within a simplified nematohydrodynamic model in which the nematic order parameter is a fixed-magnitude unit vector ${\bf n}$. The solution holds in a regime where the flow does not perturb the  nematic order imposed by the walls. 
In this study we explore systematically active flows in this confined geometry with a more general theoretical model that uses a second-rank tensor order parameter {\bf Q} to express both the magnitude and orientation of the nematic phase. The {\bf Q}-model allows for the presence of defects and biaxial, in addition to uniaxial, molecular arrangements. Our aim is to provide a unified picture, beyond the limiting regime explored previously, to serve as a guide for potential microfluidic applications that exploit the coupling between the orientational order of the molecules and the velocity field to finely control the flow and overcome the intrinsic difficulties of directing and pumping fluids at the microscale. We reveal how the nematic-flow coupling is not only dependent on geometrical constraints but also highly sensitive to material and flow parameters. We specifically stress the key role played by the activity and the flow aligning parameter and we show that solutions mostly depend on two dimensionless parameters. We find that for large values of the activity parameter the flow is suppressed for contractile particles while is either sustained or suppressed for extensile particles depending on whether they tend to align or tumble when subject to shear. We explain these distinct behaviors by an argument based on the results of the stability analysis applied to two simpler configurations: active flows confined between parallel plates with either orthogonal or perpendicular alignment at both walls. We show that the analytical laminar solution derived for the ${\bf n}$ model in the low activity limit is found also in the ${\bf Q}$ model, both analytically and numerically. This result is valid for both contractile and extensile particles and for a flow-tumbling as well as aligning nematics. We remark that this velocity profile can be derived for generic boundary conditions. To stress the more general nature of the ${\bf Q}$ model, we conclude by providing a numerical example of a biaxial three-dimensional thresholdless active flow for which we show that biaxiality is specially relevant for a weakly first-order isotropic-nematic phase transition.
\end{abstract}

\maketitle
\renewcommand{\baselinestretch}{0.75}
\section{Introduction}
Active fluids constitute a special class of complex fluids characterized by the presence of an active phase that consists of, for example, microorganisms, actomyosin networks or self-propelled colloids \cite{Saintillan18}. In these liquids the active component is able to sustain flows by continuously injecting energy at the scale of its single constituents. 

Numerous earlier studies have shown a compelling qualitative and quantitative correspondence between behaviors predicted by continuum active nematohydrodynamic models \cite{DeGennes, Thampi2016} and phenomena observed in a variety of active fluid systems \cite{Hatwalne04, Saintillan07, Wensink2012, Saw17}. Supported by this evidence, we focus on studying the behavior of active nematic liquid crystals, a class of apolar materials that display orientational order and whose particles self-propel. In the mathematical model the motility is accounted for by an active force term derived by considering that active particles can be approximated to leading order as force dipoles \cite{Thampi2016}. In these systems the transition between a passive state, in which activity is macroscopically incoherent, and an active state, characterized by a spontaneous active flow, is generally observed above a certain activity threshold \cite{Voituriez05, Aranson07, Edwards08, Sanchez12, Ravnik13}. However, there exists a family of flows that violates this rule by developing steady state velocity fields even for vanishingly small activity. 

The existence of {\it thresholdless} active flows was first reported numerically \cite{Marenduzzo07} and later formalized theoretically \cite{Green17} by identifying the asymptotic parameter regime required for their onset and the topological constraints, boundary conditions and external forcing that allow for them. A non-uniform, minimum energy  nematic profile, geometrically constrained and leading to a non-vanishing curl  active force constitutes the key ingredient  for  such a class of fluids \cite{Green17}. A realization of this situation is achieved  with  an  active nematic liquid  confined between parallel plates with hybrid anchoring at the walls: parallel on one plate, perpendicular on the other; this is one of the examples presented in \cite{Green17} and studied in \cite{Marenduzzo07} and this is the setting our study focuses on. Configurations with the same anchoring at both walls, {\it e.g.} parallel anchoring or perpendicular anchoring, lead to uniform ground states which can support a coherent unidirectional active flow only above well defined thresholds for the activity parameter as derived through the linear stability analysis \cite{Voituriez05, Edwards08}.

The motivation for studying active flows confined in a slab geometry with hybrid anchoring at the walls is of both applied and theoretical nature. On one hand this configuration is relevant to microfluidic applications, on the other, the results reported in the literature \cite{Marenduzzo07, Green17} differ and call for a more comprehensive unified picture. 

In nematic liquid crystals the coupling between the orientational order of the molecules and the flow is controlled by several material and flow parameters and the nematic configuration is highly sensitive to geometrical constraints.  The emerging complex dynamics of these active liquids is of great promise for microfluidic applications since it provides a means to control and finely tune the flow overcoming the intrinsic difficulties of directing and pumping isotropic fluids at the microscale \cite{Sengupta14, Copar20, Kos20}. Devices that direct and sort nano and micro-particles have already been presented in the literature: some exploit the anisotropic nature of the fluid to control the flow resistance and streamlines through the application of external electrical fields \cite{Na10}, some use defect lines as rails to transport colloids \cite{Ohzono12, Sengupta13} in what is referred to generically as {\it topological microfluidics}. Recently, it has been conceptualized how {\it active} liquid crystal can be exploited to design autonomous microfluidic devices \cite{Woodhouse17}. Numerical studies have also appeared to shed light on the active flow dynamics and transition from coherent to turbulent state in two or three dimensional microchannels \cite{Doostmohammadi17, Chandragiri20}.

The numerical \cite{Marenduzzo07} and theoretical \cite{Green17} studies, we will mainly refer to, are performed in two different frameworks: in \cite{Green17}  the hydrodynamic active nematic equations are expressed in terms of the director field ${\bf n}$, which represents the average long axis orientation for rod-like molecules, while in \cite{Marenduzzo07} the nematic is described by a more general tensor order parameter ${\bf Q}$ that expresses both the magnitude, $q_0$, and orientation, ${\bf n}$, of the nematic phase. The tensor order parameter formulation naturally embodies defects and allows for biaxial states \cite{Mottram14} in three-dimensions.  

The active nematic equations expressed in ${\bf Q}$ and ${\bf n}$ coincide for a uniaxial nematics with uniform $q_0$ up to second order terms in ${\bf n}$ (see Appendix C in \cite{Marenduzzo07}), yet the, respectively, numerical and analytical results reported in \cite{Marenduzzo07} and \cite{Green17} for a low activity laminar flow confined between parallel plates with hybrid alignment at the walls differ. The aim of this work is therefore two-fold: ({\it i}) provide a unifying picture for this class of active flows that generalizes the regimes explored previously and bridges the analytical results derived in \cite{Green17} with the numerical ones in \cite{Marenduzzo07} 
and ({\it ii}) take advantage of the more general nature of the ${\bf Q}$ formulation and explore the existence of biaxial thresholdless active flows. 

In this paper we first present the mathematical and numerical model we use (Sec. \ref{Qmodel}) and list a complete set of dimensionless numbers that characterize the dynamics and associated relevant regimes (Sec. \ref{DimN}). We then show, in part \ref{sec:ReviewGreen}, that the laminar flow solution derived for the ${\bf n}$ model in the low activity limit \cite{Green17} satisfies also the ${\bf Q}$ model for an appropriate choice of the free-energy parameters. More specifically, in Sec. \ref{sec:2DAnalyticalDerivation}, we find that the analytical solution found by Green {\it et al.} \cite{Green17}  can be recovered with the two-dimensional ${\bf Q}$ model and we remark that this solution can be generalized to any choice of the anchoring angle. Although no analytical expression is found for a thresholdless active flow given a three dimensional ${\bf Q}$ tensor, in Sec. \ref{sec:3DNumSolution} we show that the two-dimensional solution is a very good approximation for the three-dimensional solution found numerically. Details are also given on the parameter values required to observe such flows numerically. In Sec. \ref{2D_map_results} we compute numerically steady state solutions in a wide portion of parameter space and show that they mainly depend on two dimensionless groups: the flow aligning parameter and a number that quantifies the distance from the low activity limit and a regime where the flow does not perturb the nematic order imposed by the walls. We find that for large values of the activity the flow is suppressed for contractile particles while is either sustained or suppressed for extensile particles depending on whether they tend to align or tumble when subject to shear. We explain these distinct behaviors in \ref{IntRes} by an argument based on the results of the stability analysis applied to two simpler configurations: active flows confined between parallel plates with either orthogonal or perpendicular alignment at both walls. We find that the zero-flow solution selected dynamically by the system for a contractile nematic corresponds to a free energy stationary point that is not admitted in the ${\bf n}$ model. In Sec. \ref{GS} we compare this zero-curl stationary point with the thresholdless flow solution. Finally, to stress the more general nature of the ${\bf Q}$ model, we provide a numerical example of a biaxial three-dimensional thresholdless active flow and we show that biaxiality is specially relevant for a weakly first-order isotropic-nematic phase transition, Sec. \ref{BI} . We conclude by summarizing our findings in section \ref{Conclusions}.

\section{Mathematical and numerical model}\label{MathNum}
\subsection{The ${\bf Q}$ hydrodynamical model for active nematics}\label{Qmodel}
In the tensor order parameter model the nematic is described by a second order tensor $Q_{ij}$ that expresses both the magnitude $q_0$, and orientation ${\bf n}$, of the nematic phase. The tensor order parameter formulation naturally embodies defects and allows for biaxial states \cite{Mottram14}, in fact $Q_{ij}$ can be generically expressed as
\begin{equation}
Q_{ij} = q_0 n_i n_j - q_1 m_i m_j -(q_0+q_1)\frac{\delta_{ij}}{d},
\label{Qij}\end{equation}
where ${\bf n}$ and ${\bf m}$ are perpendicular directors of unit length that represent the axes of reflection symmetry of a biaxial nematic, $q_0$ and $q_1$ are the associated  magnitudes and $d$ is the spatial dimension of the problem. Biaxiality is possible only in 3 dimensions (3D). For a uniaxial nematic $q_1=0$ and ${\bf n}$ is an axis of rotational symmetry; in this case the order parameter reduces to
$Q_{ij} = q_0(n_in_j-\delta_{ij}/d).$

In the ${\bf Q}$-model it is customary to adopt the Landau-De Gennes free energy which consists of a distortion term multiplied by the elastic constant $K$ and bulk terms with constants $A$, $B$ and $C$ that represent the thermotropic part of the free energy \cite{DeGennes, Thampi2016, Mottram14} 
\begin{equation}
\mathcal{F} = \int d^3r \left[\frac{K}{2}(\partial_k Q_{ij})^2+\frac{A}{2}Q_{ij}Q_{ji}+\frac{B}{3}Q_{ij}Q_{jk}Q_{ki}+\frac{C}{4}(Q_{ij}Q_{ji})^2\right]. \label{FLdG}
\end{equation}
The molecular field tensor is then defined as:
\begin{align}
\mathcal{H}_{ij} & = -\frac{\delta \mathcal{F}}{\delta Q_{ij}}+\frac{\delta_{ij}}{d}Tr\frac{\delta \mathcal{F}}{\delta Q_{kl}}\nonumber \\
& = K \nabla^2 Q_{ij} - A Q_{ij}-B Q_{ik}Q_{kj}-C(Q_{lk}Q_{kl})Q_{ij}+B\frac{\delta_{ij}}{d}(Q_{lk}Q_{kl}).
\label{Hij}\end{align}
For a uniaxial nematic and $d=2$ and $3$ Eq. (\ref{Hij}) simplifies, respectively, into
\begin{align}
\mathcal{H}_{ij} &= K \nabla^2 Q_{ij}-\left(A+\frac{C}{2}q_{0}^2\right)Q_{ij}, \hspace{0.5cm} \mbox{($d=2$)} \label{Hij_2d} \\
\mathcal{H}_{ij} &= K \nabla^2 Q_{ij}-\left(A+\frac{B}{3}q_{0}+\frac{2}{3}Cq_{0}^{2}\right)Q_{ij} \hspace{0.5cm} \mbox{($d=3$)}. \label{Hij_3d}
\end{align}
The active nematic equations with $\Gamma$ as the rotational diffusivity and $\rho$ as the fluid density read
\begin{align}
\partial_i u_i & = 0, \label{incompressibility}\\
(\partial_t+u_k\partial_k)u_i & =\frac{1}{\rho}\partial_j{\Pi}_{ij},\label{NSEq}\\
(\partial_t+u_k\partial_k)Q_{ij}-S_{ij}& =\Gamma \mathcal{H}_{ij},\label{QijEq}
\end{align}
where Eq. (\ref{incompressibility}) imposes the incompressibility condition on the velocity field $u_i$, Eq. (\ref{NSEq}) is the Navier-Stokes equation with pressure term $\Pi_{ij}$, and Eq. (\ref{QijEq}) describes the evolution of the nematic tensor with $S_{ij}$ as the co-rotation term. 
The pressure term is
\begin{align}
\Pi_{ij}=&-P\delta_{ij}+2\eta E_{ij}+2\xi(Q_{ij}+\delta_{ij}/d)(Q_{kl}\mathcal{H}_{lk})\nonumber \\
&-\xi \mathcal{H}_{ik}(Q_{kj}+\delta_{kj}/d)-\xi(Q_{ik}+\delta_{ik}/d)\mathcal{H}_{kj}-\partial_iQ_{kl}(\delta \mathcal{F}/\delta \partial_jQ_{lk})+Q_{ik}\mathcal{H}_{kj}-\mathcal{H}_{ik}Q_{kj}-\alpha Q_{ij}, \label{Piij}
\end{align}
where $\alpha$ is the activity parameter. The active liquid crystal is contractile for $\alpha$ negative, and extensile otherwise. Large values of the activity parameter are expected to destabilize the nematics by triggering instabilities eventually leading to a chaotic behavior.
The co-rotation term is given by
\begin{align}
S_{ij}&=(\xi E_{ik}+\Omega_{ik})(Q_{kj}+\delta_{kj}/d)+(Q_{ik}+\delta_{ik}/d)(\xi E_{kj}-\Omega_{kj})-2\xi(Q_{ij}+\delta_{ij}/d)(Q_{kl}\partial_ku_l)\label{Sij},
\end{align}
 where $E_{ik}$ and $\Omega_{ik}$ are respectively the symmetric and antisymmetric part of the velocity gradient tensor, that is the strain rate tensor and the vorticity tensor, while the parameter $\xi$ is the flow-aligning parameter. The co-rotation term expresses the response of the nematic field to the extensional and rotational part of the velocity gradients, a low value of the flow-aligning parameter induces tumbling of the particles while larger values correspond to a flow-aligning tendency. The range of $\xi$ values that correspond to a flow-tumbling and flow-aligning behavior can be found in analogy with the $\bf{n}$ model: when $\lambda = \xi\frac{2+q_0 d - 2q_0}{q_0 d}$ is larger than unity particles are in the flow-aligning regime. In the case of a biaxial nematics the flow-tumbling and flow-aligning distinction will still hold true but the additional $q_1$ parameter expressing the magnitude of biaxiality will enter into the expression for $\lambda$: $\lambda = \xi (2+q_0d-2q_0-2q_1)/(q_0 d).$
  
The active nematohydrodynamic equations (\ref{incompressibility})-(\ref{QijEq}) are solved numerically using a hybrid Lattice Boltzmann (LB) finite-difference method \cite{Vincenzi15}. More precisely, the nematic pressure term and the equation for the evolution of the $Q_{ij}$ tensor are integrated through a second order finite-difference scheme. The time integration of $Q_{ij}$ is performed by means of an explicit second order Adams-Bashforth time stepping scheme. The contribution of the active and passive nematic pressure terms is added to the Navier-Stokes equation as an external forcing. The Navier-Stokes equations are then integrated through the Lattice Boltzmann method  \cite{Succi}. The LB method makes the code ideally suited for parallel computing, the code is parallelized on CPUs with an MPI distributed parallelism.

For this study the equations are integrated in a channel that extends from $y=0$ to $y=L$ with no-slip boundary conditions and hybrid anchoring at the walls. Specifically, for most of our calculations, the nematic order parameter is aligned parallel to the wall at $y=0$ and perpendicular to it at $y=L$, that is, for $\theta = \arctan(n_y/n_x)$ with $n_x$ and $n_y$  the $x$ and $y$-components of the director field ${\bf n}$, one has $\theta(y=0)=0$ and  $\theta(y=L)=\pi/2$. Different anchoring angles have been considered in Sec. \ref{GS}. See Fig. \ref{rms_dev} ({\it left}) for a schematic representation of the geometrical configuration. 
We carry out the numerical integration on 1D-domains. This implies that only the $x$-component of the velocity is non-zero and only the $y$-derivatives of the velocity and order tensor fields are non-zero, hence instabilities can only manifest and grow in the $y$-direction. The order parameter $Q_{ij}$ is allowed to have non-zero components on either a 2D plane or in the 3D space, that is, $Q_{ij}$ can be either two-dimensional ($d=2$) or three-dimensional ($d=3$). The analytical solutions in Sec. \ref{sec:2DAnalyticalDerivation} are derived for a two-dimensional $Q_{ij}$, the numerical results reported in Sec. \ref{sec:3DNumSolution} are obtained for both a three-dimensional and two-dimensional $Q_{ij}$, while the numerical results shown in Sec. \ref{2D_map_results} and Sec. \ref{GS} are for a three-dimensional $Q_{ij}$.
\subsection{Dimensionless parameters}\label{DimN}

Several dimensional parameters appear in eq. (\ref{NSEq}) and (\ref{QijEq}): $\eta$, $\alpha$, $\rho$, $\Gamma$, $K$, $A$, $B$, $C$. 
Three characteristic length scales can be identified in this model: ({\it i}) a length scale representing the core size of topological defects, $l_c$, ({\it ii}) an active length marking the scale at which active energy is injected into the system \cite{Giomi15, Joanny20}, $l_a$, and ({\it iii}) a geometrical length scale, $L$, representing the width of the channel. 
The scale of the defect core, $l_c$, is estimated through a Taylor series expansion around the minimum of the free energy eq. (\ref{FLdG}). For a three-dimensional nematic tensor this yields 
$$l_c=\sqrt{\frac{K}{A/3+2Bq_{0,eq}/9+2Cq_{0,eq}^2/3}},$$ where $q_{0,eq}$ is the equilibrium value of the magnitude of the nematic tensor for a uniform and undistorted nematic. The active length scale is estimated balancing the active and passive nematic terms $l_a=\sqrt{K/|\alpha|}$.
  
These three characteristic length scales combined with the characteristic velocity scale of the flow, $v_{0}$, and the dimensional parameters that do not appear in the definition of $l_c$ and $l_a$ provide the following complete set of dimensionless parameters
 
\begin{enumerate}
\item The balance between the inertia and viscous terms in eq. (\ref{NSEq}) gives the Reynolds number $Re = \rho v_0 L/\eta, $ note, however, that we are concerned with steady state solutions and effectively one-dimensional profiles for which the material derivatives in both eq. (\ref{NSEq}) and (\ref{QijEq}) are zero. The Reynolds number is therefore always zero and not relevant to the problem under consideration. 

\item A balance between the viscous terms and the passive nematic terms in eq. (\ref{NSEq}) yields the Ericksen number $Er = \eta v_0 L/K.$ 

\item A balance between the active terms and the passive nematic terms gives the ratio between the active length scale and the system characteristic length scale: $\Pi_1 = \alpha L^2/K= \sign(\alpha)L^2/l_a^2.$ 

\item The ratio between the characteristic length of the defect core and the channel length scale provides $\Pi_2 = l_c/L.$

\item Finally, the dimensionless number used to identify the {\it frozen director limit} (FDL)  discussed in the following sections is $\Pi_3 = \Gamma \eta.$
\end{enumerate}

A sixth dimensionless parameter that appears in the model in dimensionless form is the flow aligning parameter $\xi$. Numerically, it is necessary to resolve all the relevant length scales, particularly the defect core, $l_c$, and the active length, $l_a.$ 
The time scale $\tau = L^2/K \Gamma$ provides a useful reference on the relaxation time scale and the duration of the initial transient that precedes convergence to a steady state solution. 

Out of the six dimensionless parameters we expect our system to be independent of $Re$, as explained above, $\Pi_2$, since we select system sizes much larger than the characteristic defect core ($\Pi_2 \ll 1$), and $Er$ since in the absence of an external forcing the characteristic velocity $v_0$ depends on the other model parameters. The Ericksen number will coincide with $\Pi_1$ when the viscous and active forces balance ($v_0\propto\alpha L/\eta$), with $\Pi_3$ when the molecular field term and the co-rotation term balance in Eq. (\ref{QijEq}) ($v_0\propto\Gamma K/L$), and will be a function of $\Pi_1$ and $\Pi_3$ in all the other cases. In conclusion we expect our problem to depend on three independent parameters: $\Pi_1$, $\Pi_3$ and $\xi$.

\section{Results} 
\begin{figure} \centering
\resizebox{8cm}{!}{\includegraphics{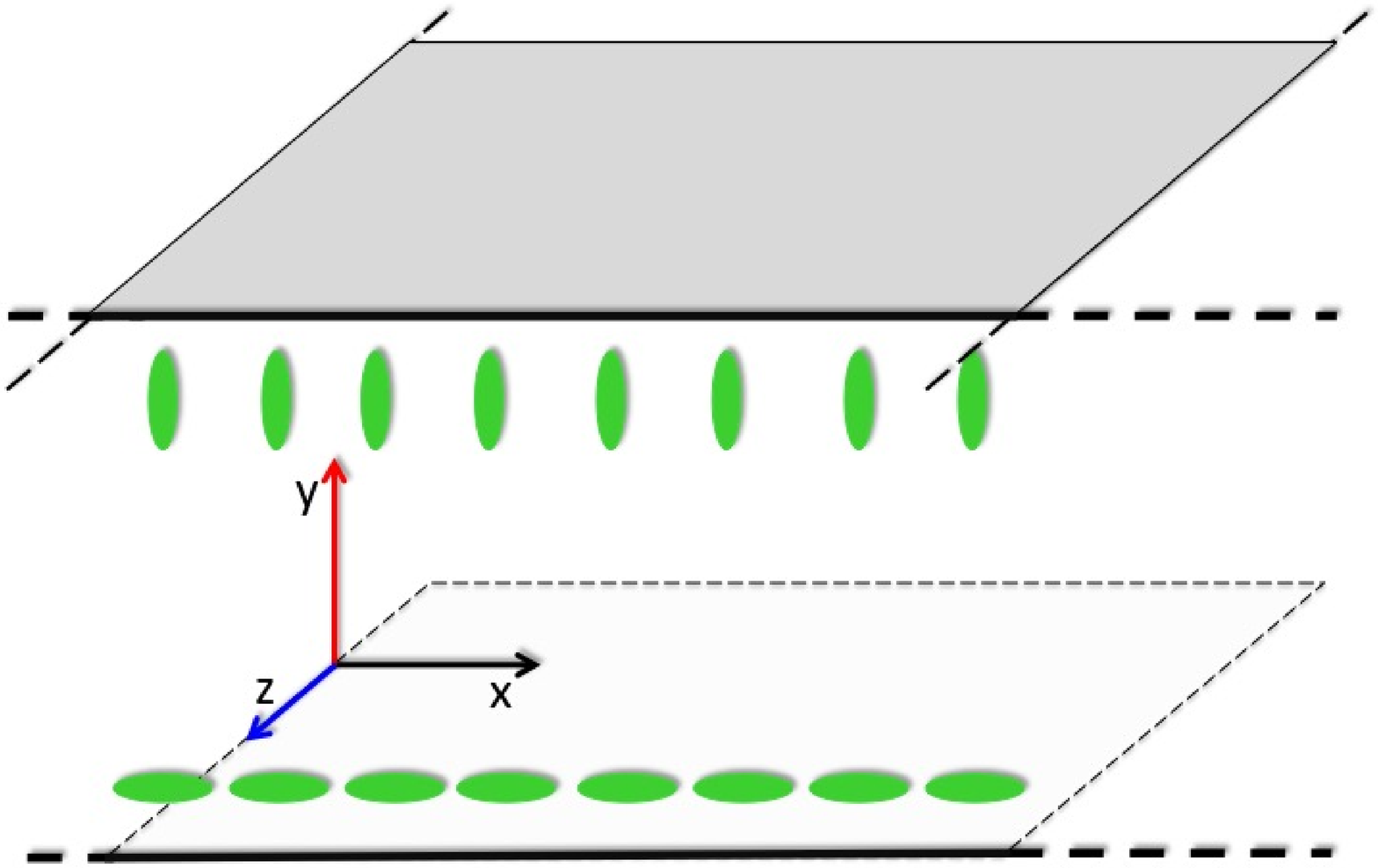}}
\resizebox{8cm}{!}{\includegraphics{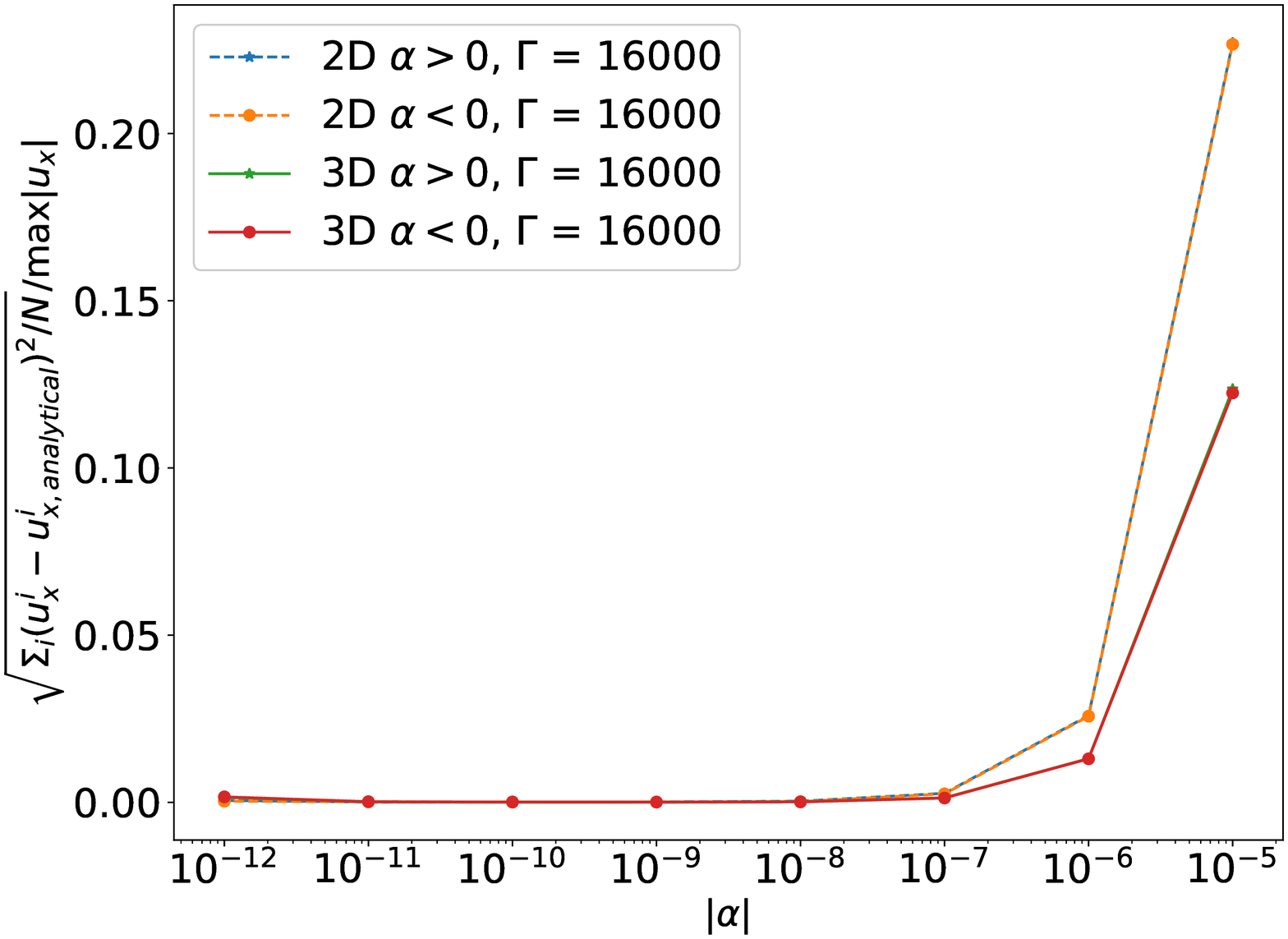}}
\caption{\label{rms_dev}
{\it Left:} schematic representation of a channel with hybrid alignment at the walls. The channel walls located at $y=0$ and $y=L$ extend to infinity in the $x$ and $z$ directions. The anchoring of the active nematic liquid crystals is parallel to the $y=0$ wall (homogeneous anchoring) and normal to the $y=L$ wall (homeotropic anchoring).  The numerical integration is performed in 1D. {\it Right:} Normalized root mean square error measuring the deviation of the numerical velocity profile from the analytical expression Eq. (\ref{Vel_analytical}) as a function of the magnitude of the activity parameter. In the formula reported on the $y$-axes $N$ is the number of grid-points. The numerical solution is obtained by integrating the full active nematohydrodynamic equations with either a two-dimensional or three-dimensional tensor order parameter $Q_{ij}$.The parameters of the simulations are  $\nu = 0.33$,  $\rho=2$, $L=256$, $t=500000$,  $\xi=0.7$, $\Gamma = 16000$, $K=5\cdot 10^{-6}$. For the two-dimensional case $q_0=0.9998$, $A=-2.5\cdot 10^{-6}$, $B = 0 $, $C= 5 \cdot 10^{-6}$, for the three-dimensional case $q_0=0.5$, $A=0$, $B = -C= -3\cdot 10^{-5}$. These parameter values correspond to: $10^{-2}<|\Pi_1|<10^{5}$, $\Pi_2\approx6.8\cdot10^{-3}$, $\Pi_3 = 10560$.} 
\end{figure}

\subsection{Thresholdless active flow in a two-dimensional channel with mixed boundary conditions}\label{sec:ReviewGreen}
\subsubsection{Analytical solutions in the {\bf n} and ${\bf Q}$ model}\label{sec:2DAnalyticalDerivation}
As Green {\it et al.} \cite{Green17} noted, in steady state and in the absence of fluid flow the equation for the evolution of the director field in the ${\bf n}$-model simply reduces to the Euler-Lagrange equation for minimizing the free energy with constraint $|{\bf n}|=1$:
$\frac{\delta F}{\delta n_i}-\left(\frac{\delta F}{\delta n_j}n_j\right)n_i = 0,$
where $F$ is the Frank free energy. 
If the director field is in the ground state it is shown that
the velocity field is zero only if the pressure gradient balances the active force term $f_{a,i}=\partial_j(n_j n_i)$  exactly \cite{Green17}. Hence, a sufficient condition for the onset of thresholdless active flows is that the active force has a non-vanishing curl \cite{Green17}. Under this condition and in the regime where the nematic  is not distorted by the flow, referred to as the FDL ($\Pi_3 \gg 1$), analytic expressions for the flow field can be derived. Green {\it et al.} \cite{Green17} provide some solutions for various geometrical configurations, among them, a two-dimensional channel flow with hybrid alignment at the walls as shown in Fig. \ref{rms_dev} ({\it left}).
 
In a 2D channel with walls at $y=0$ and $y=L$ and mixed boundary conditions: $n_x(x, 0)=1$, $n_y(x, 0)=0$, $n_x(x, L)=0$, $n_y(x, L)=1$, the equilibrium nematic profile 
\begin{equation} n_x = \cos\left(\frac{\pi y}{2 L}\right), \: n_y = \sin\left(\frac{\pi y}{2 L}\right) \label{analytical_nx_ny}
 \end{equation}
 induces an active force with non-vanishing curl \cite{Green17}.
In the zero activity limit and Stokes flow regime the velocity field can be computed analytically (see Appendix G in \cite{Green17}) leading to 
\begin{equation}
u_x = -\frac{\alpha L}{2 \pi \eta}\left(\cos\frac{\pi y}{L}+2\frac{y}{L}-1\right).\label{Vel_analytical}
\end{equation}

Finding a solution analogous to (\ref{analytical_nx_ny})-(\ref{Vel_analytical}) in the $\bf{Q}$-model requires solving $\mathcal{H}_{ij} = 0$ in two-dimensions with mixed boundary conditions: $n_x(x, 0)=1$, $n_y(x, 0)=0$, $n_x(x, L)=0$, $n_y(x, L)=1$ and the assumption of uniform $q_0$. Given that $Q_{ij}$ is a function of $y$ only and $|\bf{n}|$ is unitary, we have

\begin{equation}
\frac{Q_{xx}}{q_{0}} = \frac{1}{2}-n_y^2, \hspace{0.5cm}    
\frac{Q_{xy}}{q_{0}} = \frac{Q_{yx}}{q_{0}} = n_y\sqrt{1-n_y^2}, \hspace{0.5cm}
\frac{Q_{yy}}{q_{0}} = n_y^2-\frac{1}{2}.         
\end{equation}
Since $q_0 $ is uniform, Eq. (\ref{Hij_2d}) can be rewritten as $\mathcal{H}_{ij} = K \nabla^2 Q_{ij}-a Q_{ij}$ where $a$ is a constant and the stationary point condition $\mathcal{H}_{ij}=0$ corresponds to the system of ODEs
 
 \begin{align}
 -2 (n_y'^{2} + n_y n_y'') &= \bar{a} \left(\frac{1}{2}-n_y^2\right), \label{ODE1}\\
 \left[ \frac{-3n_y n_y'^2+2n_y^3n_y'^2+n_y''-3n_y^2n_y''+2n_y^4n_y''}{(1-n_y^2)^{3/2}}\right] & = \bar{a} n_y\sqrt{1-n_y^2}, \label{ODE2}\\
  2 (n_y'^{2} + n_y n_y'') &= \bar{a} \left(n_y^2-\frac{1}{2}\right), \label{ODE3}
 \end{align}
for respectively the $xx$, $xy$ and $yy$ component of the molecular field. Here $n_y'$ and $n_y''$ are, respectively, the first and second total derivative of $n_y$, while $\bar{a} = a/K$. Note that the first and third equation coincide. If we replace 
 $n_y''=- n_y'^2/n_y + \bar{a} (n_y^2 - 0.5)/(2 n_y)$ obtained from Eq. (\ref{ODE1}), into Eq. (\ref{ODE2}), we get
 $n_y'^2=\bar{a}(n_y^2-1)/4$,
which, solved with the mixed boundary conditions gives 
 \begin{equation}
 n_y(y) = -\frac{1}{2} i e^{- i \frac{\pi y}{2L}}(1-e^{i\frac{\pi y}{L}}) = \sin\left(\frac{\pi y}{2 L}\right).
 \label{VitelliSol2D}\end{equation}
From Eq. (\ref{VitelliSol2D}) we have that $a=-K\pi^2/L^2$, hence for small $K$ and large $L$ ({\it e.g.} the values we have chosen for our numerical calculations: $K=5\cdot 10^{-6}$ and $100\le L\le256$) one has $a \ll 1$. This shows that the ground state configuration found in \cite{Green17} [Eq. (\ref{analytical_nx_ny})] for mixed boundary conditions and the active nematic equations expressed in terms of the director field ${\bf n}$, satisfies also the active nematic equations formulated in terms of the tensor order parameter provided that $a$ is non-zero and as given above. The value of $a$ prescribes the values for the constants $A$, $C$ and $q_0$ according to expression (\ref{Hij_2d}).

To find the velocity profile that corresponds to the nematic profile (\ref{VitelliSol2D}), we substitute it into expression (\ref{Piij}) and solve Eq. (\ref{NSEq}). We then have: \begin{align}
\Pi_{ij}=&-P\delta_{ij}+2\eta E_{ij}-K(\partial_i Q_{kl}\partial_jQ_{lk})-\alpha Q_{ij},
\end{align}
the third term in $\Pi_{ij}$ is non-zero only for $i=j=y$, it is constant, and hence does not contribute to Eq. (\ref{NSEq}) that, as in \cite{Green17}, reduce to
\begin{align}
\eta u''_x-\alpha q_0 (n_y n'_x+n_x n'_y)&=0,\\
-P'-\alpha q_0 (n_y^2)'=0,
\end{align}
which, once solved with no-slip boundary conditions gives the same solution as in \cite{Green17}, here Eq. (\ref{Vel_analytical}), except for an extra multiplicative factor $q_0$. 

Note that eq. (\ref{VitelliSol2D}) and (\ref{Vel_analytical}) are just a special case of a broader family of solutions with anchoring conditions $\theta(y=0)=\theta_0$ and $\theta(y=L)=\theta_L$. By defining $\Delta \theta = \theta_L-\theta_0$ we have that the general solution is:
\begin{align}
n_x &= \cos\left(\frac{\Delta \theta y}{L}+\theta_0\right), \: n_y = \sin\left(\frac{\Delta \theta y}{L}+\theta_0\right),\\
u_x &= -\frac{\alpha L q_0}{4 \eta \Delta \theta}\left\{\cos\left(\frac{2\Delta\theta y}{L}+2\theta_0\right)-  \frac{y}{L}\left[\cos(2\Delta\theta+2\theta_0)-\cos(2\theta_0)\right]-\cos(2\theta_0)\right\},\label{GenVitSol}
\end{align}
the case $\Delta\theta=0$ corresponds to the degenerate case with uniform $n_x$ and $n_y$ and zero velocity. 
Consider also that for a 1D geometry, the zero-curl condition for the active force is satisfied whenever the off-diagonal terms of $Q_{ij}$ are zero.

The analytical solutions (\ref{analytical_nx_ny}) and (\ref{Vel_analytical}) derived for a 2D $Q_{ij}$ in a 1D-geometry cannot be easily extended to the case of a 3D $Q_{ij}$. In fact in 1D only the trivial $q_0=0$ solution satisfies the system $\mathcal H_{ij}=0$, for $\mathcal H_{ij}$ as in Eq. (\ref{Hij_3d}), mixed boundary conditions, and the simplifying assumptions of a uniaxial nematic, uniform $q_0$ and constant $n_z$. Similarly, no analytical solutions were found for the less restrictive conditions of a uniaxial nematic and ({\it i}) uniform $q_0$ and variable director field ${n_z}(y)$ or ({\it ii}) constant $n_z$ and variable $q_0(y)$. The stationary point solution for $\mathcal H_{ij}=0$ with a 3D $Q_{ij}$ can be found numerically and will simultaneously involve a non-homogeneous $q_0$, a variable director field, and biaxiality. This is shown in the next section where we also stress that for our choice of parameters the deviations from uniform $q_0$ and uniaxiality are small.
\subsubsection{Numerical analysis of the thresholdless active flow}\label{sec:3DNumSolution}

Our first aim is to verify solution (\ref{Vel_analytical}) numerically for a 2D and a 3D ${\bf Q}$-tensor. This velocity profile is found in the limit of small activity, $|\Pi_1|\ll 1,$ and a `frozen' nematic, $2q_0/\Gamma\ll\eta,$ or, for $q_0\approx constant$ and of order one $\Pi_3\gg1.$ Reproducing (\ref{Vel_analytical}) numerically requires a careful selection of the model parameters because deviations from its perfectly symmetric shape are significant even for small values of the coupled passive nematic terms, expression (\ref{Piij}), and co-rotation terms, Eq. (\ref{Sij}). The following considerations guided us in identifying the right parameter range to replicate  (\ref{Vel_analytical}): a stable numerical solution of the diffusion terms ($\partial_t Q_{ij} = \Gamma K \nabla^2Q_{ij}$) in Eq. (\ref{QijEq}) for a central difference second order Adams-Bashforth time stepping scheme requires $\Gamma K<2/21$, hence, the large values of $\Gamma$ called for by the FDL require correspondingly small values of $K$ and force even smaller $\alpha$ to satisfy the small activity limit. 

Figure \ref{rms_dev} ({\it right}) shows the normalized root mean square (RMS) deviation of the numerical results from the analytical solution as a function of the magnitude of the activity parameter $|\alpha|$. In this plot $\Pi_3 = 10560,$ while $10^{-2}<|\Pi_1|<10^{5}$. The numerical solution is in excellent agreement with the analytical one in the small activity limit and deviates from it as $|\alpha|$ increases. As expected, the deviation from solution (\ref{Vel_analytical}) is continuous with the model parameters. In quantitative terms we find that the RMS error is below 0.26\% for $|\alpha|\le10^{-7}$ (or $|\Pi_1|=10^3$) suggesting that in reality the condition for small activity, $\Pi_1\ll 1$ holds for a wider range than predicted.
We also find that for a 3D order parameter the deviation of the minimum-energy solution from Eq. (\ref{analytical_nx_ny}) is small and involves a variation of $q_0$ in proximity of the walls as well as a small degree of biaxiality far from the boundaries. These features have been verified numerically by letting 
\begin{equation}\partial_t Q_{ij} = \Gamma  \mathcal{H}_{ij}\label{relaxation}\end{equation}
relax to equilibrium for a 3D ${\bf Q}$. For $|\alpha|=10^{-12}$ the 3D ${\bf n}$-profile shows a deviation of $\approx 0.015\%$ from the analytical profile (\ref{VitelliSol2D}), while the variation of $q_0$, as well as the degree of biaxiality estimated as the difference between the two lowest eigenvalues are approximately $\approx 0.01$\%. Therefore we conclude that the 2D ${\bf Q}$ solution is a very good approximation for the 3D ${\bf Q}$ case.

In closing, retrieving the analytical solution  (\ref{Vel_analytical}) numerically served the double purpose of testing the code and proving that the parameter regimes where the solution exists can be accessed and explored numerically.

\subsection{Transition from symmetric to asymmetric velocity profiles}\label{2D_map_results}

\begin{figure} \centering
\resizebox{14cm}{!}{\includegraphics{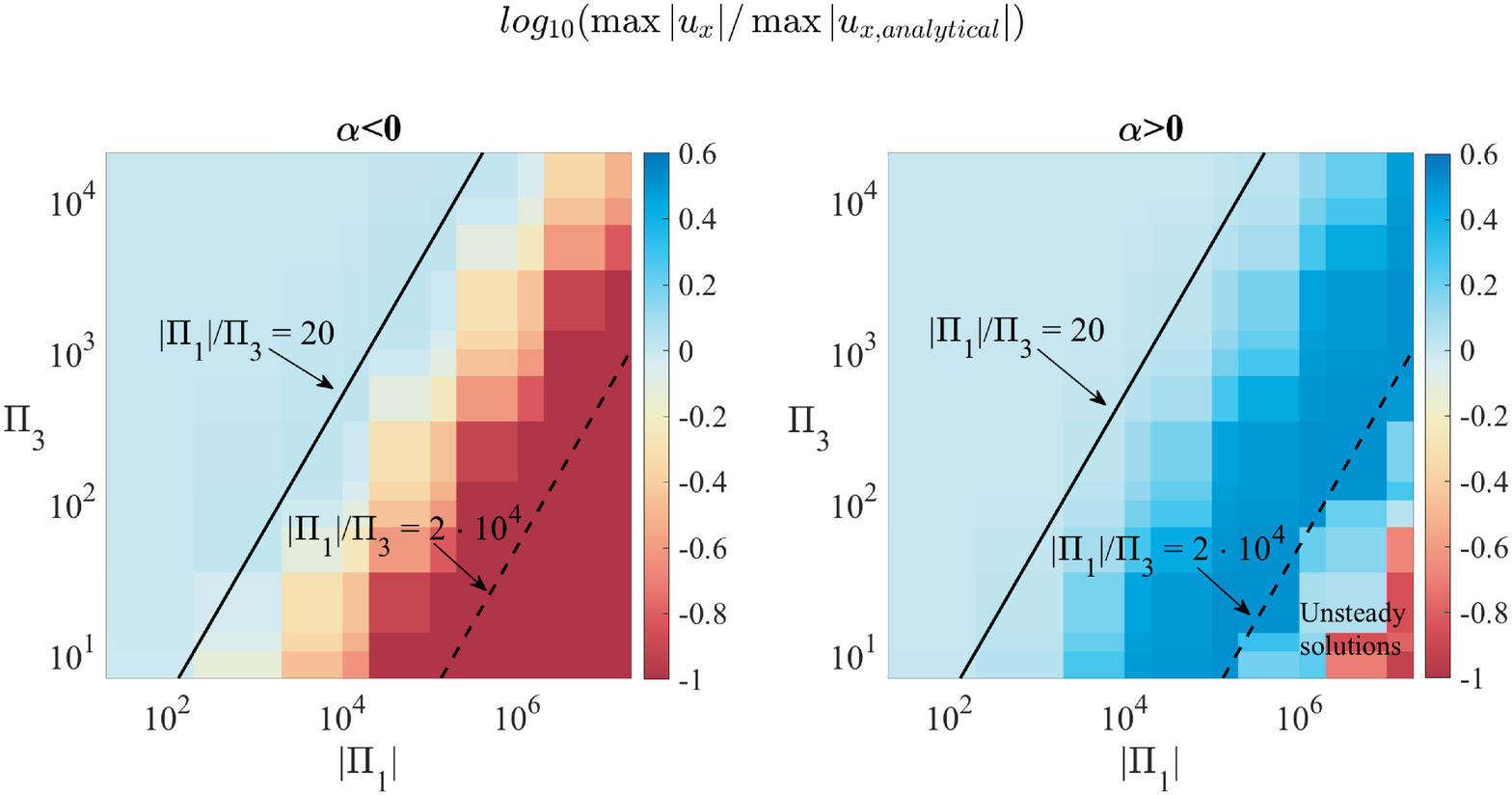}}
\put(-380,160){(a)}
\put(-200,160){(b)}\\
\resizebox{7cm}{!}{\includegraphics{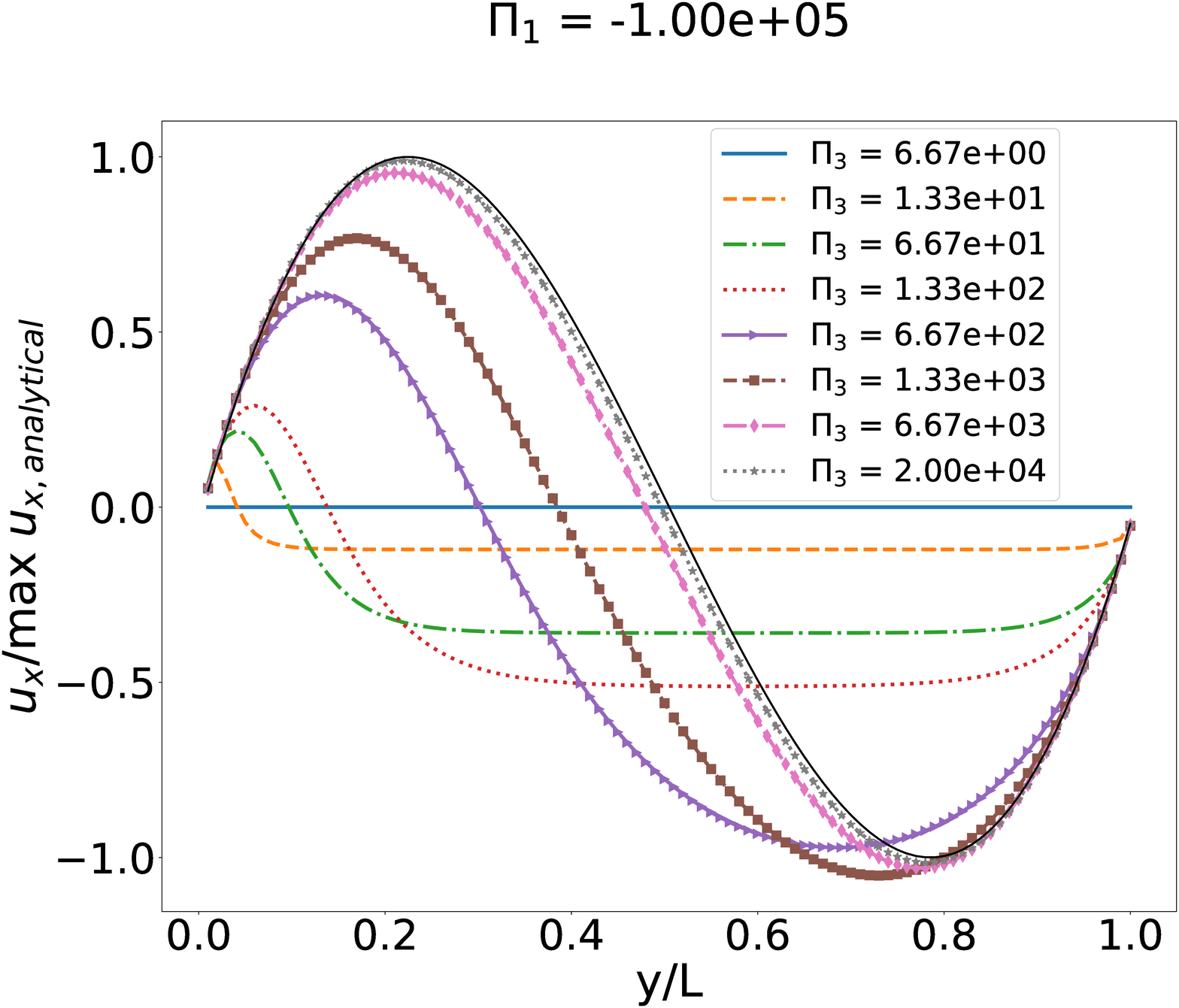}}
\resizebox{7cm}{!}{\includegraphics{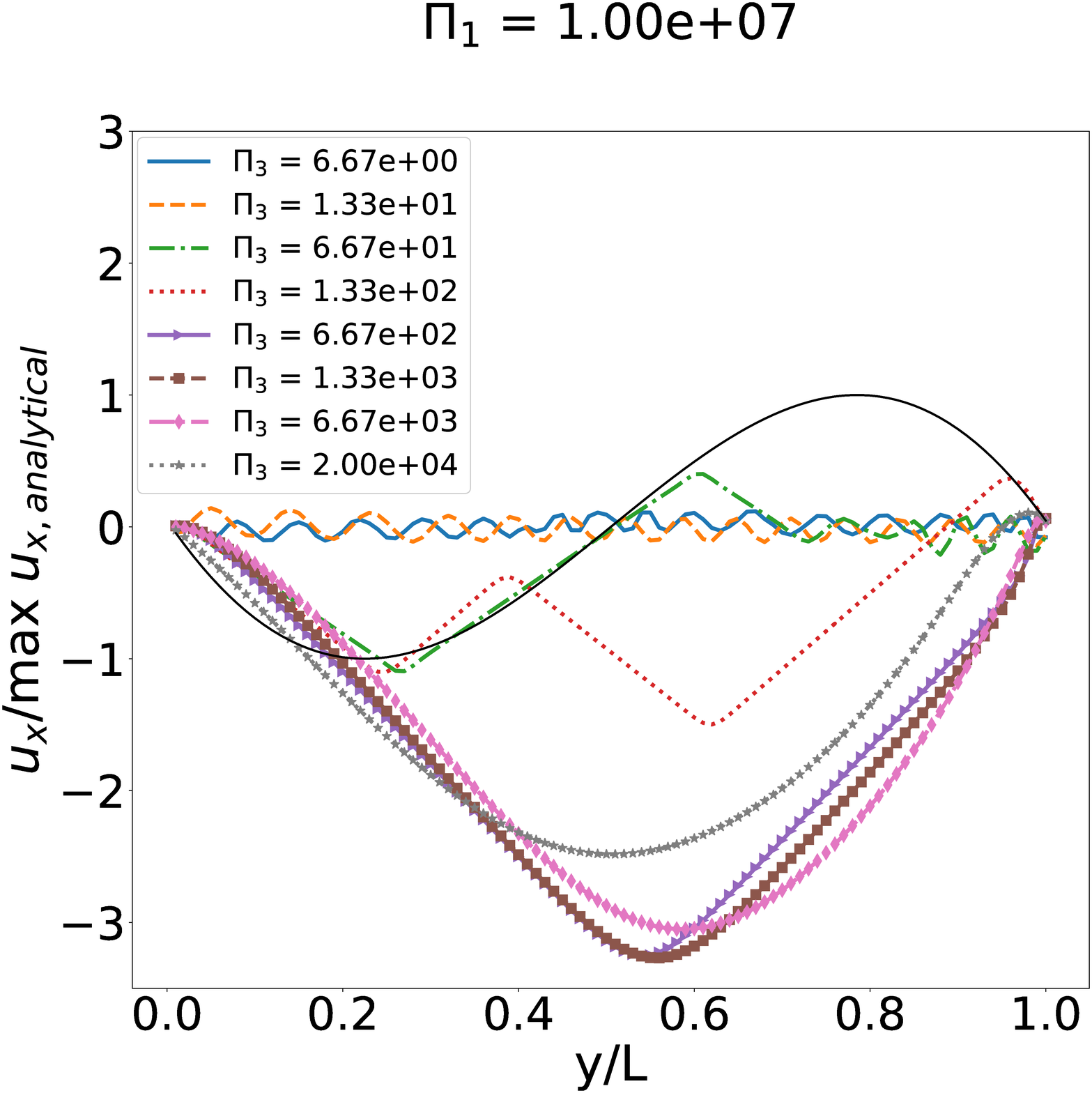}}
\put(-400,140){(c)}
\put(-200,140){(d)}\\

\begin{minipage}{0.45\textwidth}
\centering
\resizebox{7.2cm}{!}{\includegraphics{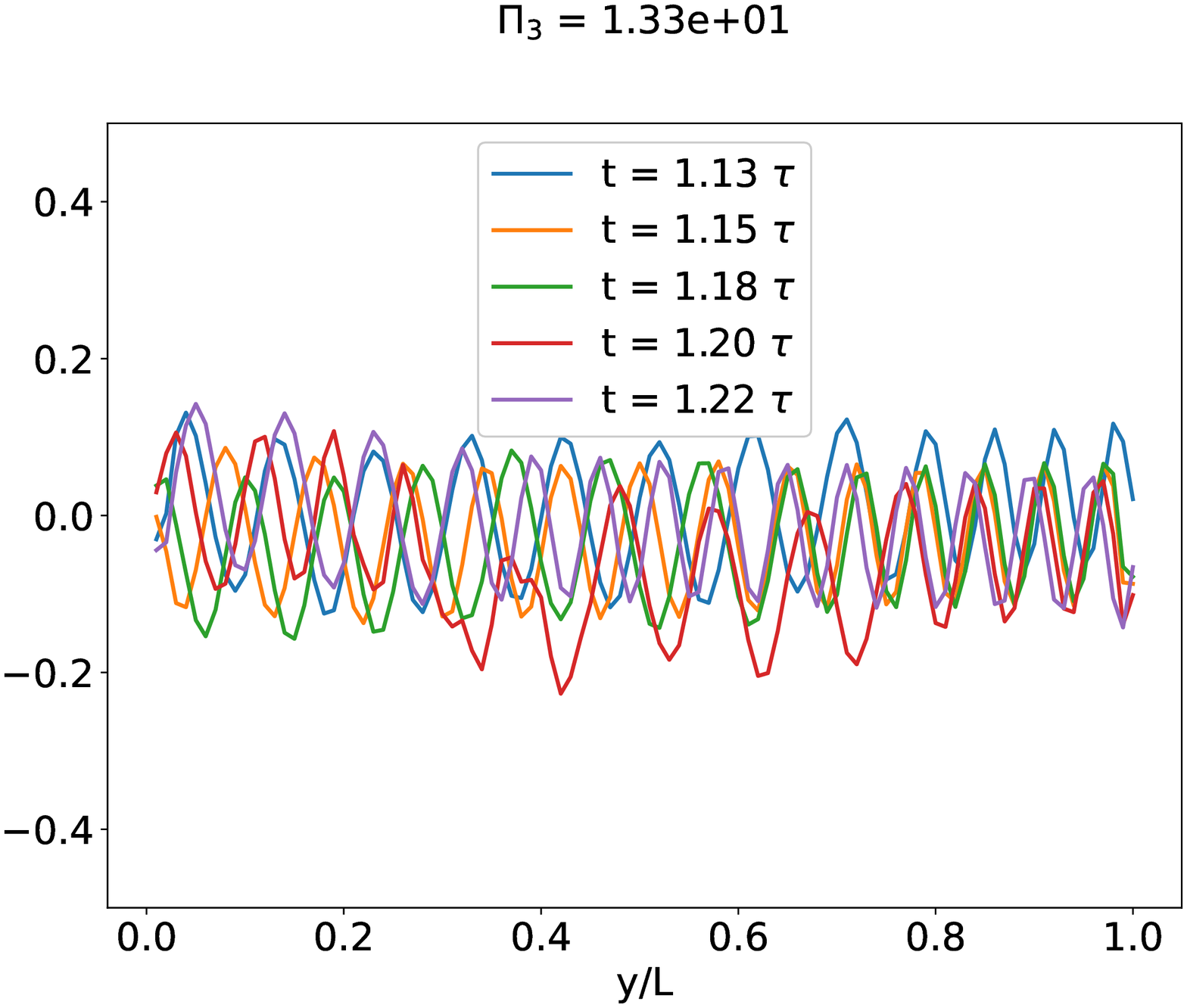}}
\put(-200,140){(e)}\\
\end{minipage}%
\begin{minipage}{0.45\textwidth}
\centering
\resizebox{7.5cm}{!}{\includegraphics{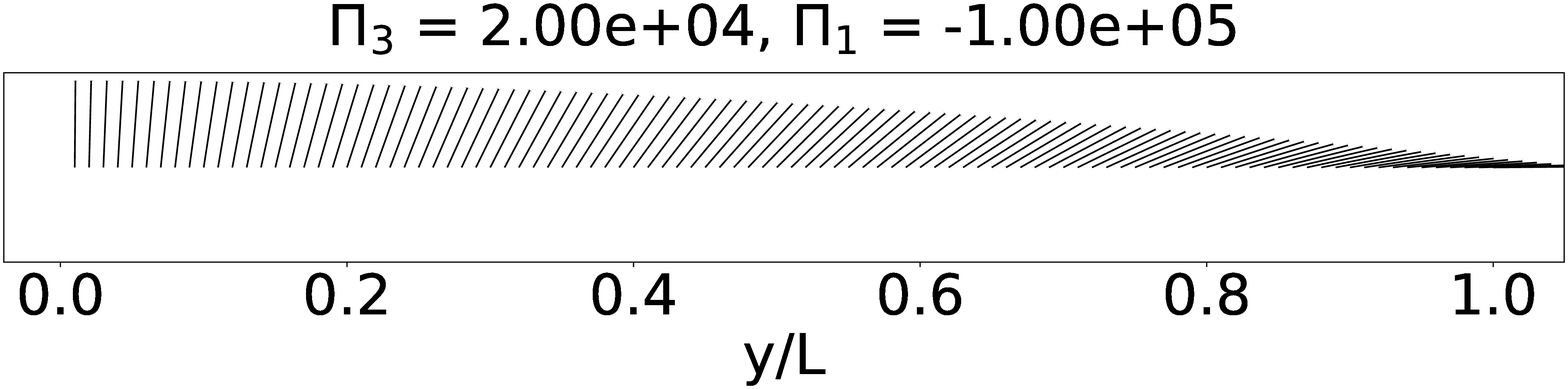}}
\put(-220,30){(f)}\\
\resizebox{7.5cm}{!}{\includegraphics{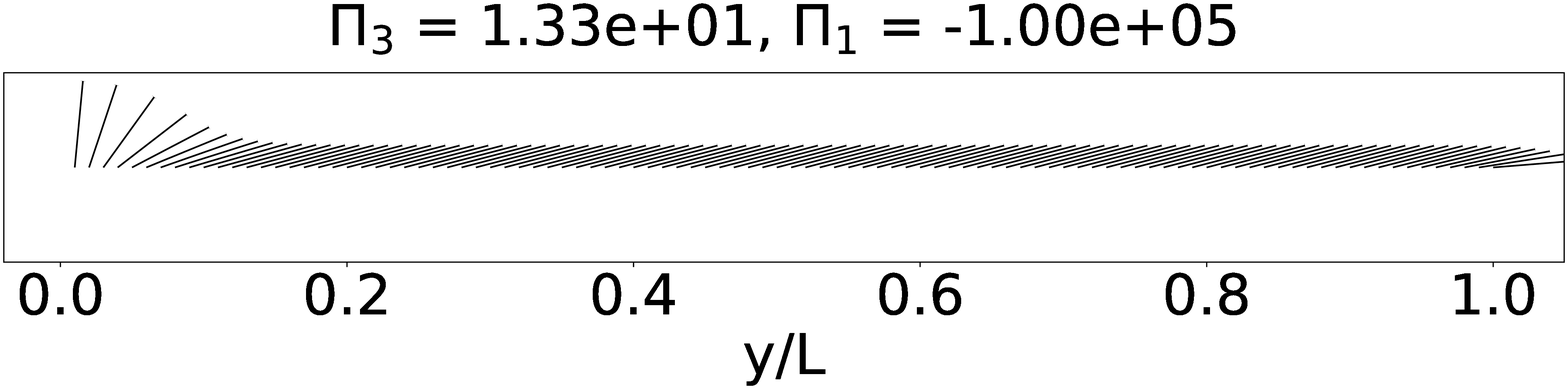}}
\put(-220,30){(g)}\\
\resizebox{7.5cm}{!}{\includegraphics{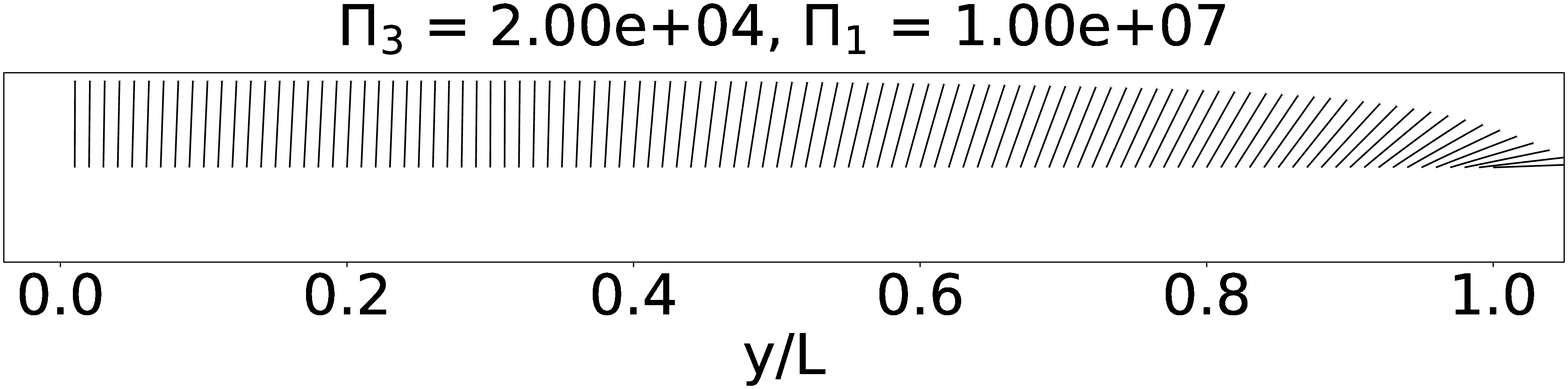}}
\put(-220,30){(h)}\\
\end{minipage}%

%
%
%
\caption{\label{2D_map}
{\it Top}: Base ten logarithm of the maximum magnitude of the velocity computed numerically and rescaled by the maximum of the analytical profile (\ref{Vel_analytical}) for a contractile (a) and extensile (b) activity parameter, $\alpha$. This normalized velocity is plotted as a function of $\Pi_3$ and $|\Pi_1|$ for a flow aligning nematics ($\xi = 0.7$). The axis are in logarithmic scale and the map reports results for a total of 330 separate calculations. Simulations with smaller values of $\Pi_3$ are more demanding in computational terms given the slower convergence: for our choice of parameters the slowest calculations run for $9.8 \cdot 10^8$ time steps. {\it Middle}: rescaled velocity profiles corresponding to cases that lie on a vertical cut of the colormaps in (a) and (b) as specified by the legend and title of the plot for a negative (c) and positive (d) value of the activity parameter. In (d) the flow profiles with $\Pi_3<6.67 \cdot 10^2$ are unsteady, in these cases we display the configuration at the final time $T_{fin}$. We label as unsteady those calculations for which the RMS deviation in the last ten saved time-steps spaced by approximately $ \tau/50$ time units is below 0.25\%. The thick black curves correspond to the analytical solution (\ref{Vel_analytical}). In (e) we show some intermediate configurations for a selected unsteady case.} {\it Bottom:} (f)-(h) director field orientation associated to three cases as detailed by the plot titles. 
\end{figure}

\begin{figure} \centering
\resizebox{8cm}{!}{\includegraphics{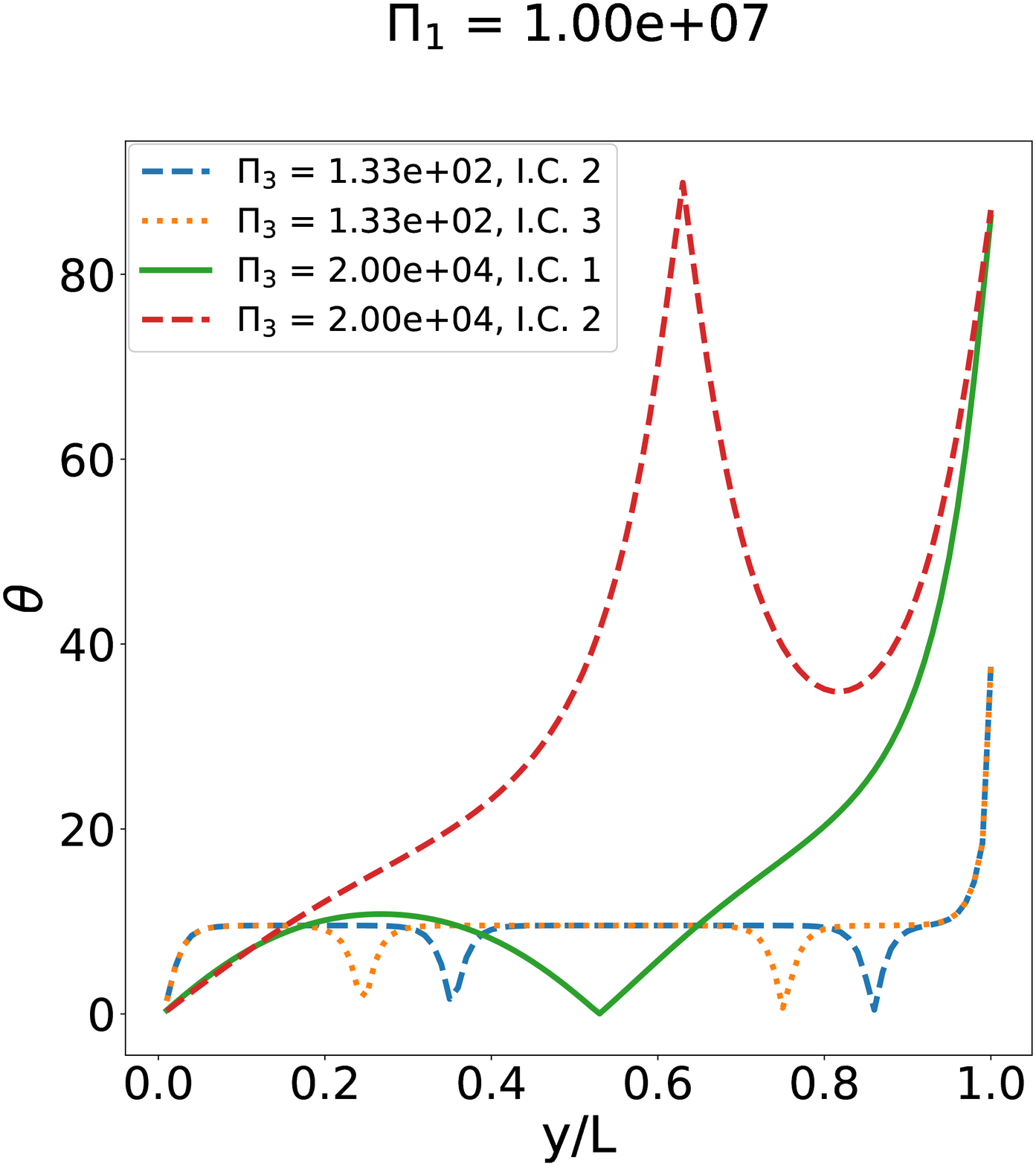}}
\resizebox{8cm}{!}{\includegraphics{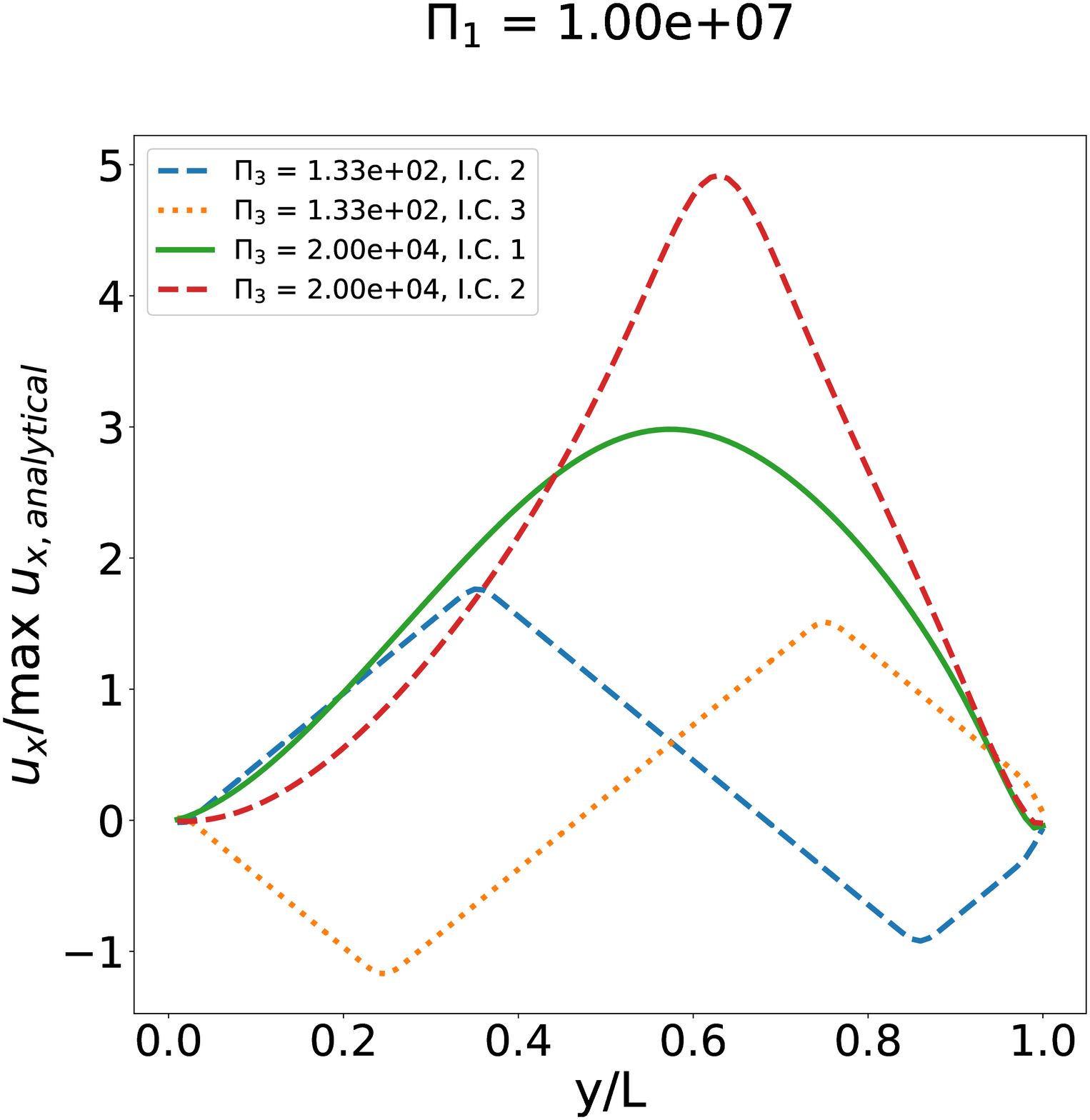}}
\put(-440,170){(a)}
\put(-220,170){(b)}\\
\caption{\label{bistability} Nematic angle $\theta = \arctan(n_y/n_x)$ (a), and velocity profile (b), for two of the calculations reported in Fig. \ref{2D_map}(b), (d) and additional simulations performed for the same values of the model parameters and different initial conditions (I.C.). In the legend `I.C. 1' corresponds to the initial condition of Fig. \ref{2D_map}, that is, an $n_x=1$ field perturbed by random noise, `I.C. 2' is given by $n_x = \cos(\pi y/ L), \: n_y = \sin(\pi y/ L)$, while `I.C. 3'  is Eq. (\ref{analytical_nx_ny}). In all cases the velocity field is initialized to zero.}
\end{figure}

\begin{figure} \centering
\resizebox{16cm}{!}{\includegraphics{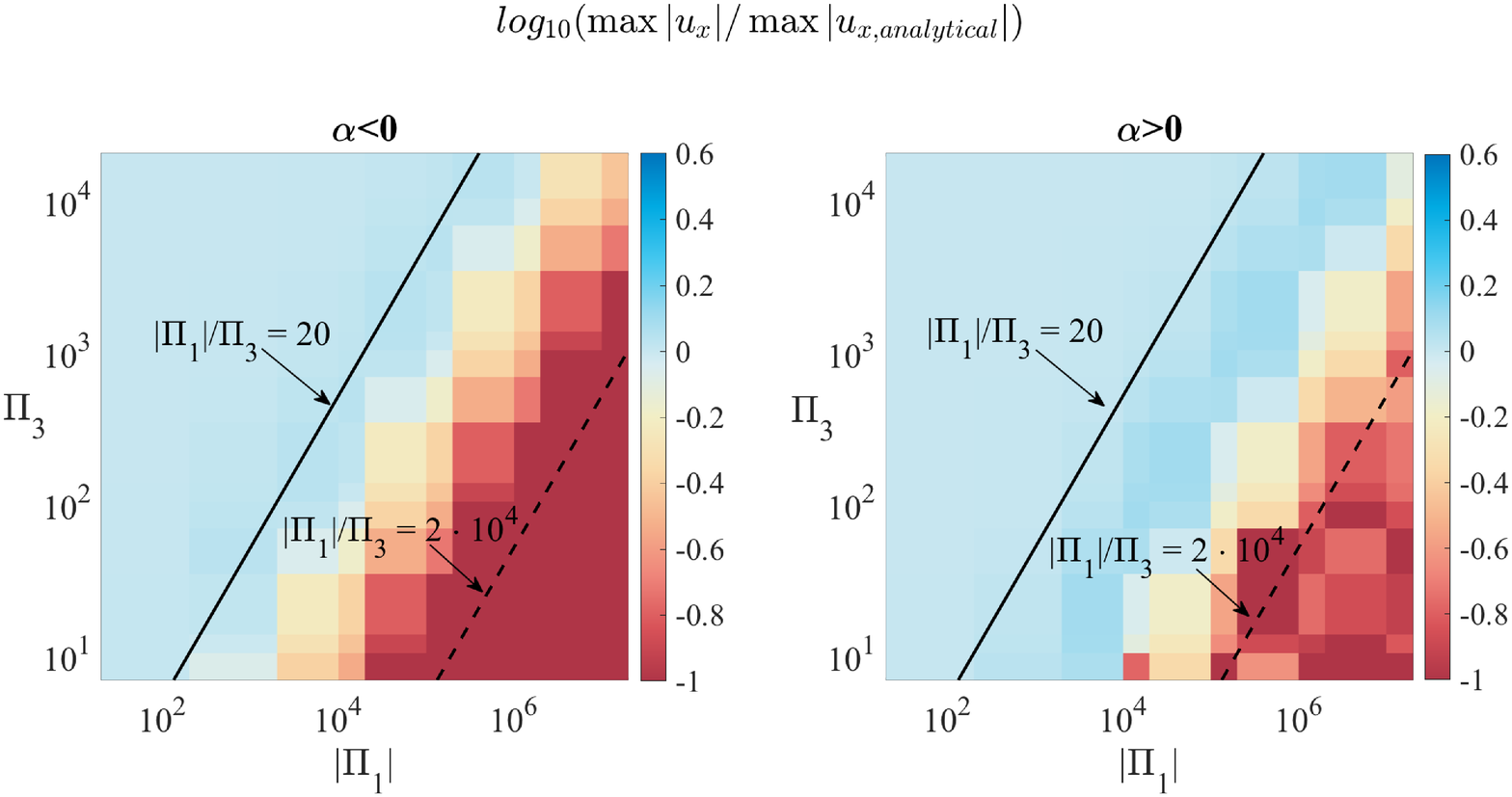}}
\put(-430,180){(a)}
\put(-230,180){(b)}\\
\resizebox{7cm}{!}{\includegraphics{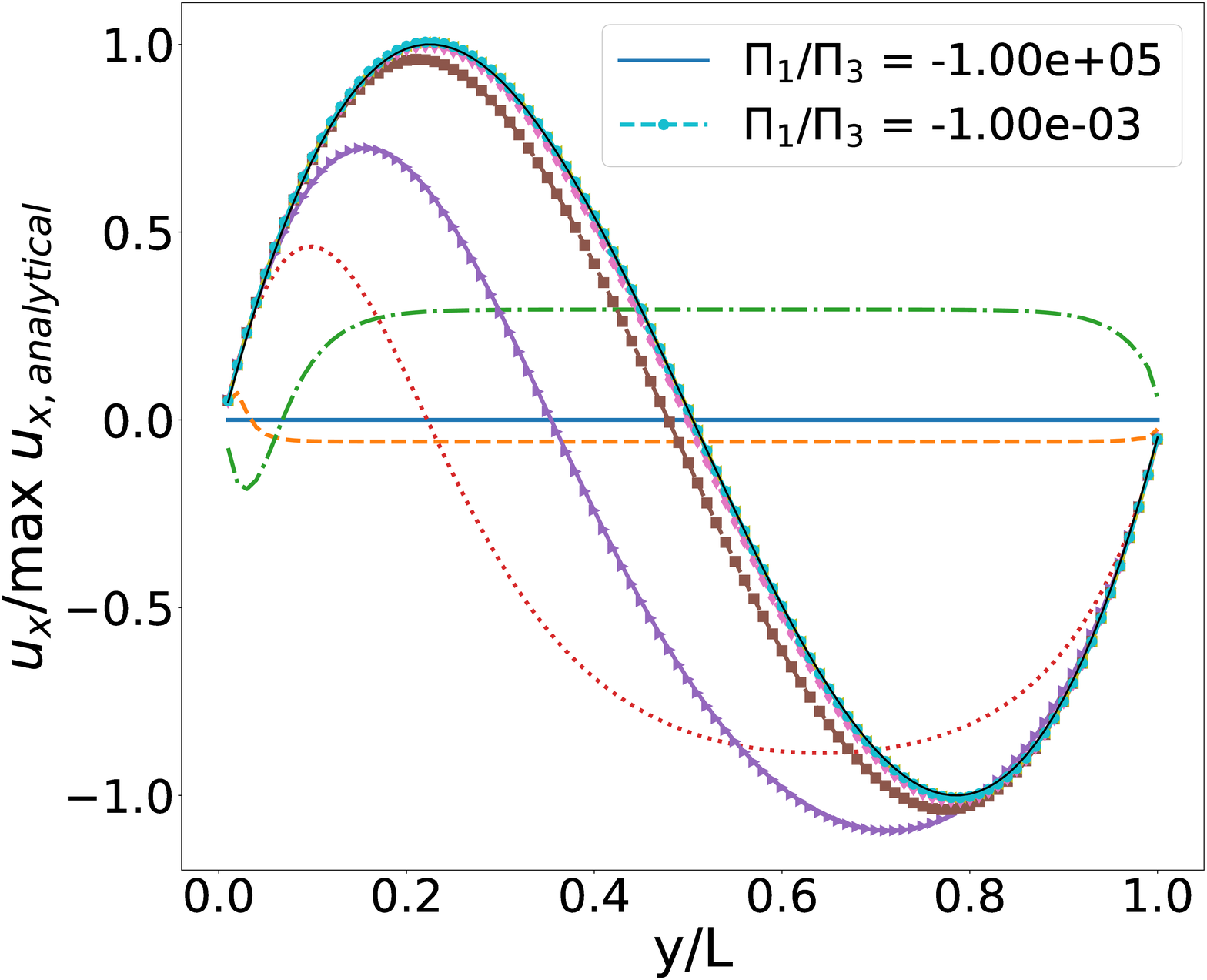}}
\resizebox{7cm}{!}{\includegraphics{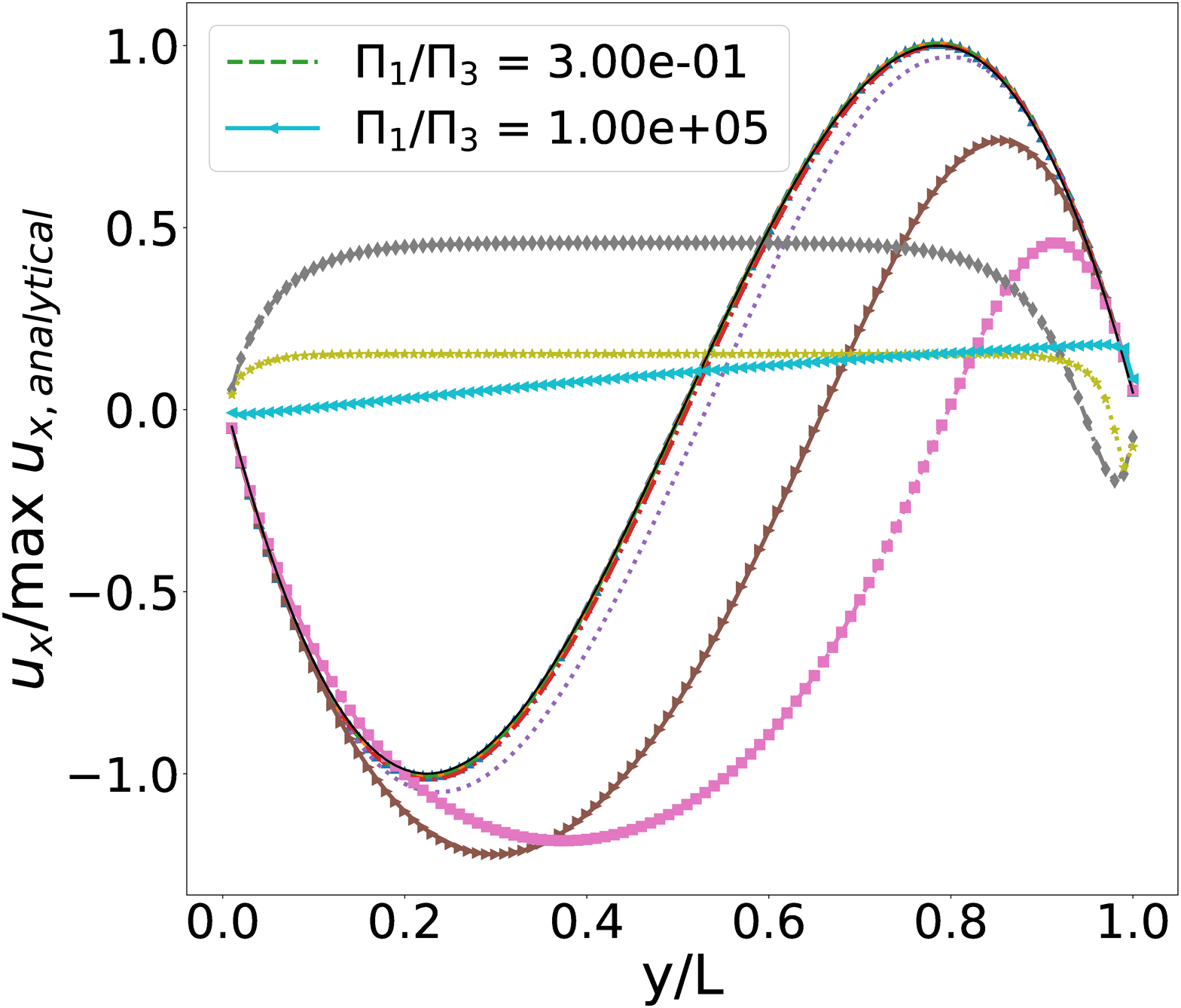}}
\put(-400,130){(c)}
\put(-200,130){(d)}\\
\resizebox{7cm}{!}{\includegraphics{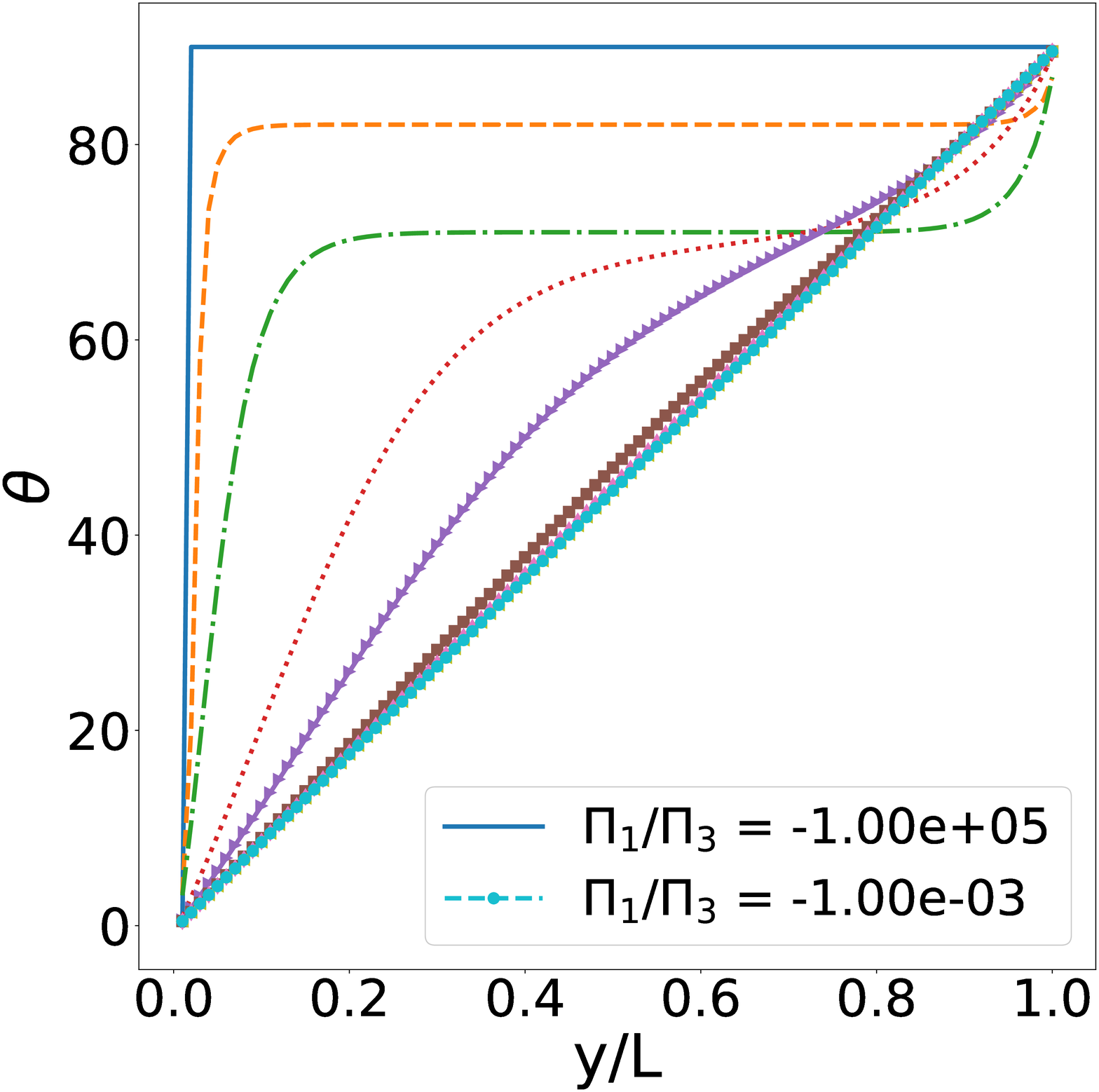}}
\put(-180,130){(e)}
\caption{\label{2D_map_tumbling}
{\footnotesize {\it Top}: Base ten logarithm of the maximum magnitude of the velocity computed numerically and rescaled by the maximum of the analytical profile (\ref{Vel_analytical}) for negative (a) and positive (b) values of the activity parameter and a flow-tumbling nematics ($\xi = 0.3$). The axis correspond to the $\Pi_3$ and $|\Pi_1|$ parameters and are in logarithmic scale. Two $|\Pi_1|/\Pi_3$ isolines  are shown in panel (a)-(b): $|\Pi_1|/\Pi_3=2 \cdot 10^{4}$ (dashed-black line) and $|\Pi_1|/\Pi_3=20$ (solid-black line). {\it Middle}: rescaled velocity profiles corresponding to calculations that lie on a diagonal cut of the colormaps in (a) and (b), the cuts originate at the top-left corner of the maps and run perpendicular to the $|\Pi_1|/\Pi_3$ isolines. For clarity the legend only labels the two curves that corresponds to the extreme values of $\Pi_3/ \Pi_1$. The thick black lines correspond to the analytical solution (\ref{Vel_analytical}). Observe the remarkable resemblance of the rescaled velocity profiles in (c) with those reported in Fig. \ref{2D_map} (c) for the flow aligning case. {\it Bottom}: nematic angle $\theta = \arctan(n_y/n_x)$ corresponding to the rescaled velocity profiles in (c). The analytical solution corresponds to a straight line, while the zero-flow solution correspond to a discontinuous profile that suddenly jumps close to the bottom wall from $n_x=1$ to $n_y=1$, this is allowed in the ${\bf Q}$-model by a concomitant $q_0 = 0$. }}
\end{figure}

\begin{figure} \centering
\resizebox{8cm}{!}{\includegraphics{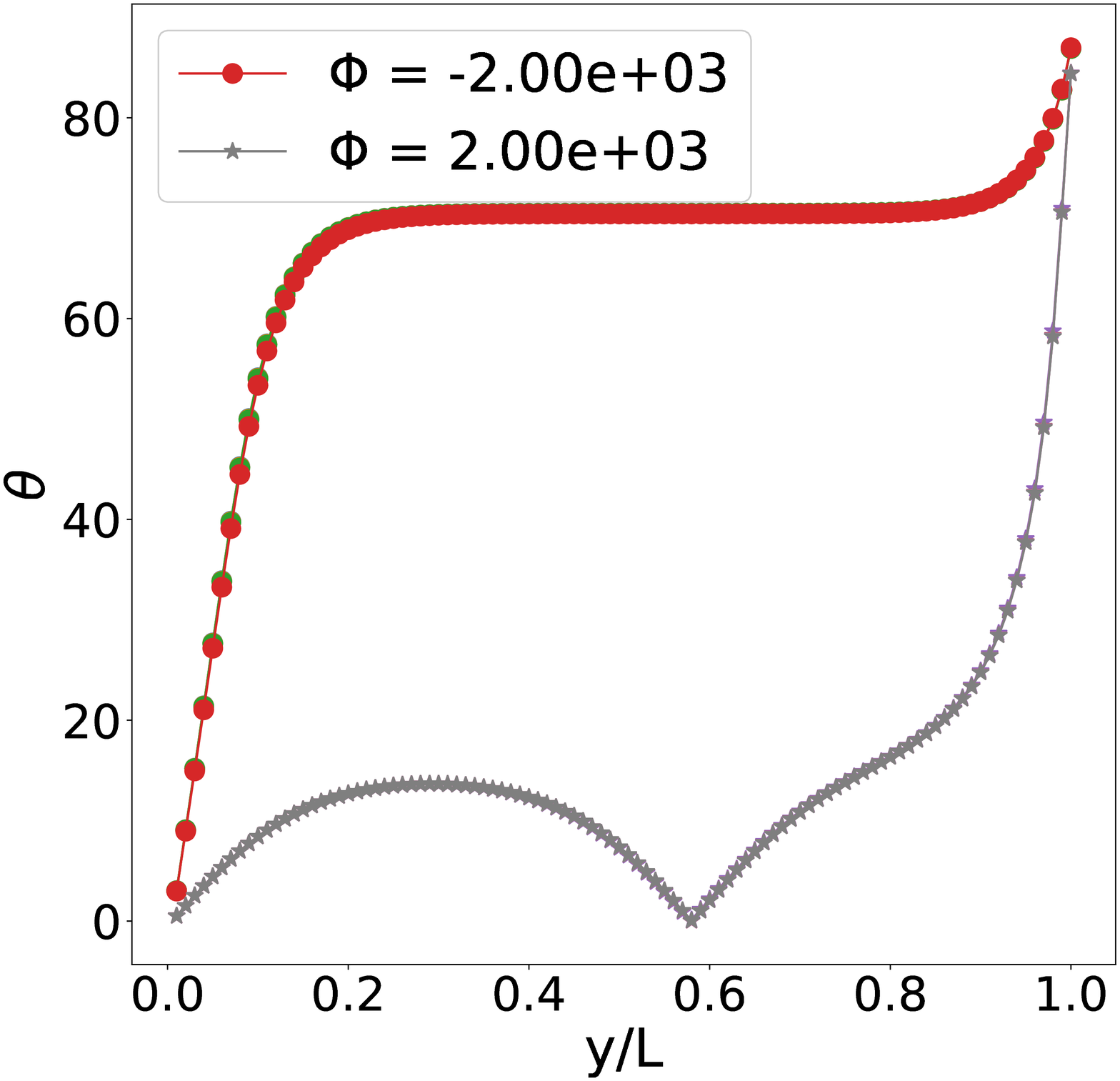}}
\resizebox{8cm}{!}{\includegraphics{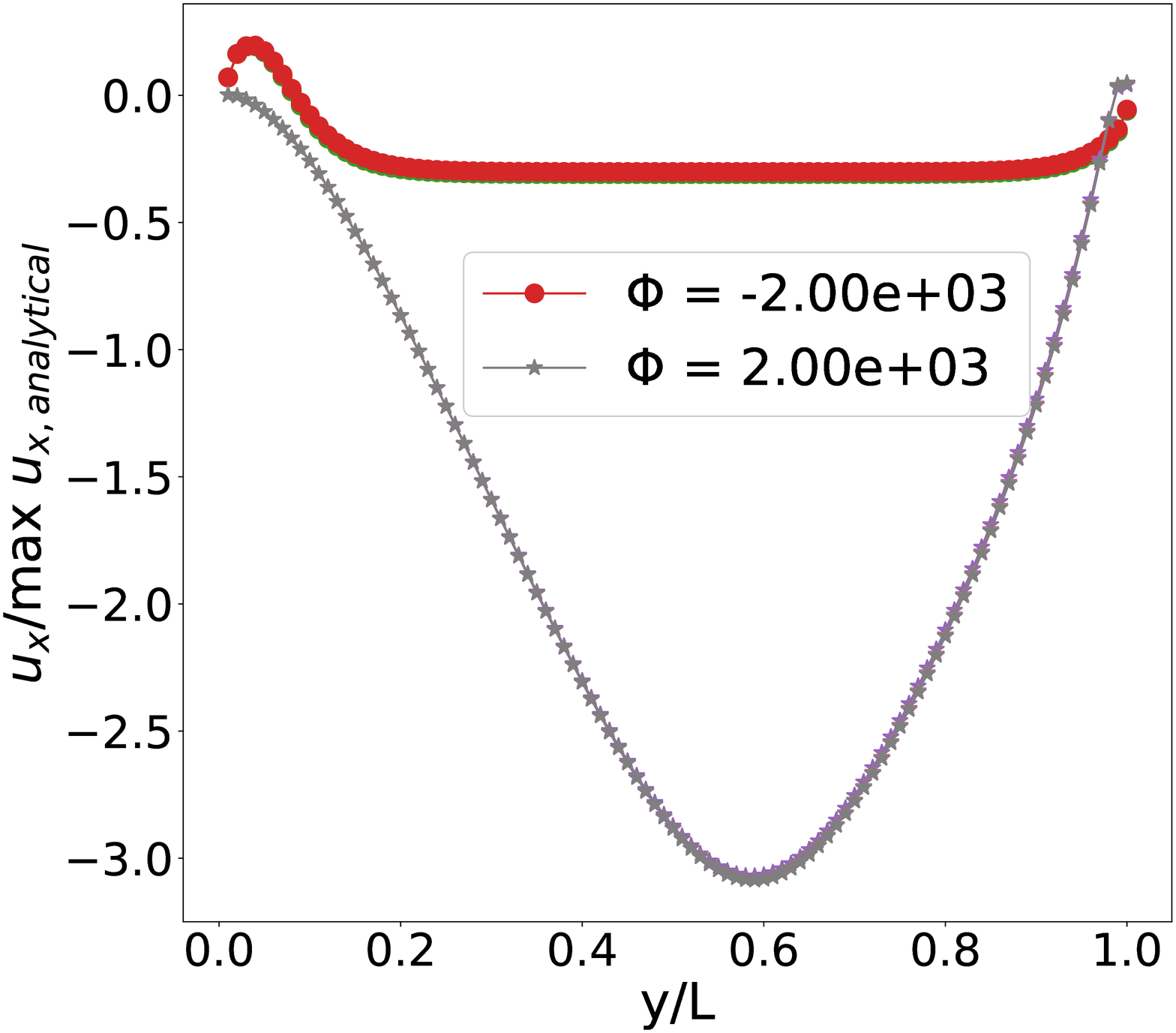}}
\put(-440,170){(a)}
\put(-220,170){(b)}\\
\resizebox{8cm}{!}{\includegraphics{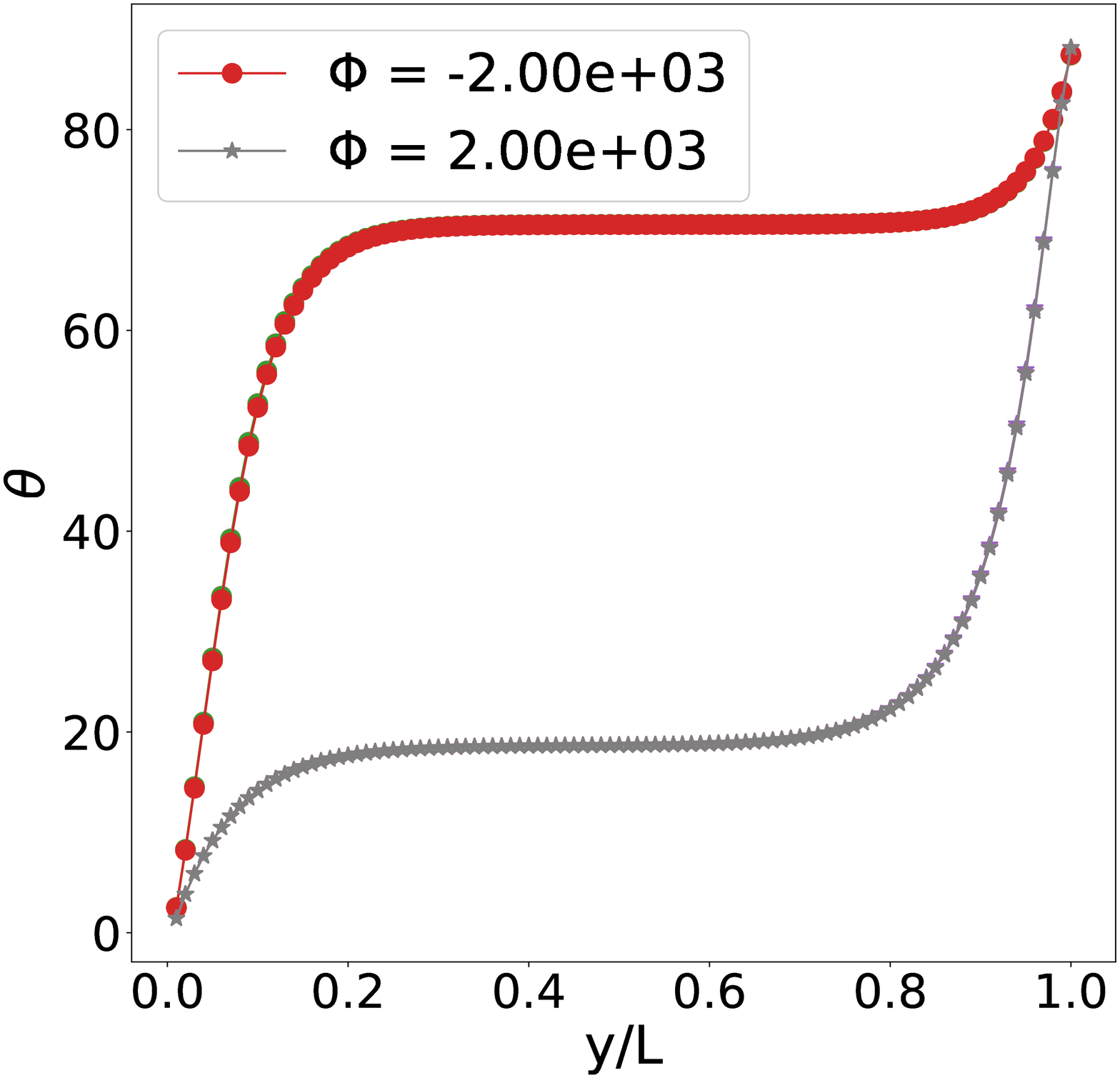}}
\resizebox{8cm}{!}{\includegraphics{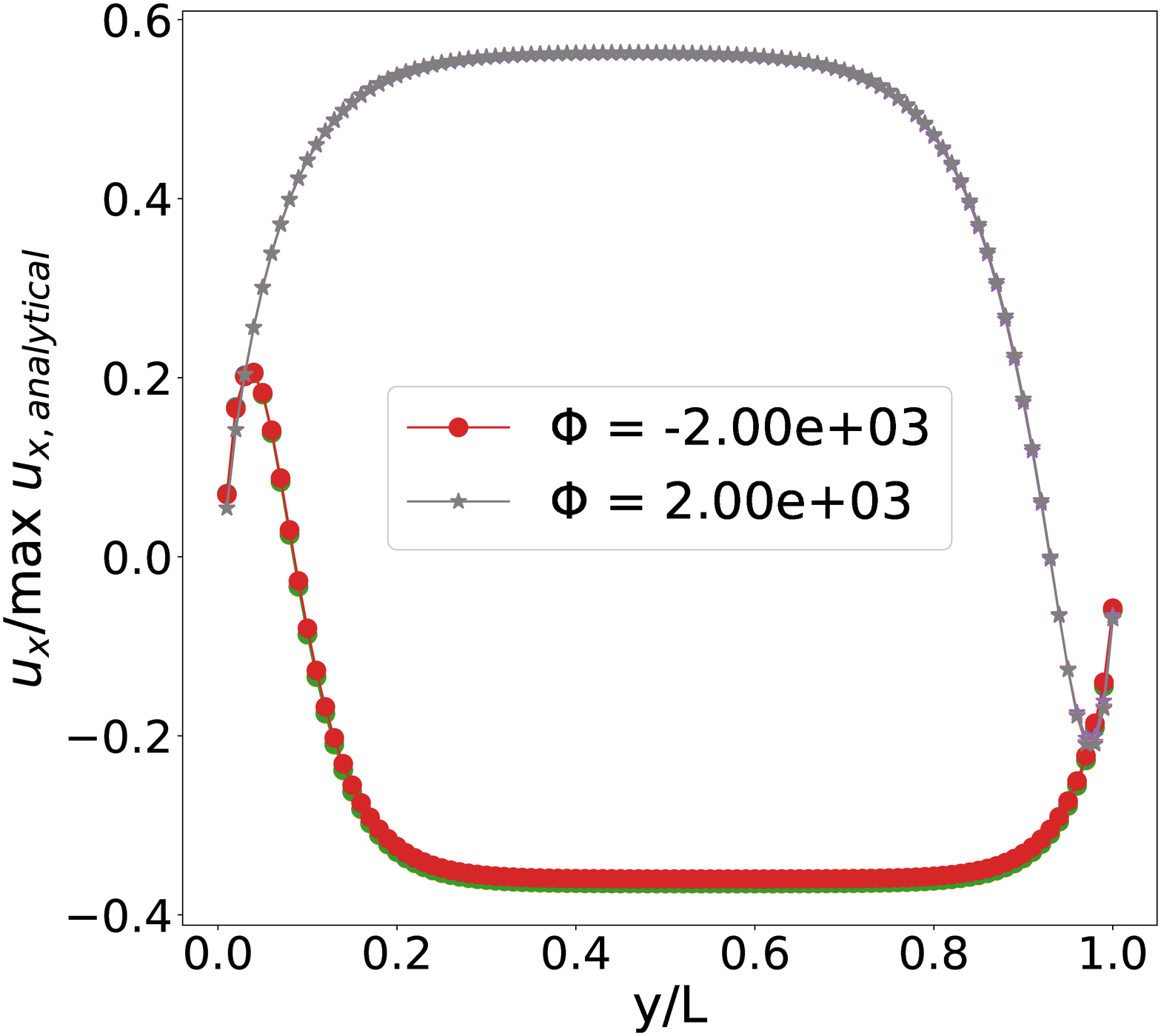}}
\put(-440,170){(c)}
\put(-220,170){(d)}\\
\caption{\label{collapsed_prof}
Nematic angle $\theta = \arctan(n_y/n_x)$ (a), (c)  and velocity profile (b), (d) rescaled by  the maximum of the analytical profile for the flow aligning case (a)-(b) and the flow tumbling case (c)-(d) and two different values of the $\Phi=\Pi_3/ \Pi_1$ dimensionless group.}
\end{figure}

The velocity profile in the FDL is anti-symmetric with respect to the mid-point of the channel. This symmetry is readily broken by increasing $\alpha$ (or in dimensionless terms $|\Pi_1|$) and/or decreasing $\Gamma$ (or $\Pi_3$) as a consequence of the fact that the passive nematic terms in expression (\ref{Piij}) as well as term (\ref{Sij}) start playing a role. The interplay between these terms and the active term also breaks the positive/negative $\alpha$-symmetry embodied in solution (\ref{Vel_analytical}): in general, the behavior for negative $\alpha$ differs from that of positive $\alpha$. An example is provided by the low-activity asymmetric velocity profiles reported in Fig. 13-15 in \cite{Marenduzzo07}. 

In this section we explore how the velocity and nematic profiles evolve in parameter space moving away from the FDL regime. The aim is to expand on previous studies and provide a unified picture that includes both the laminar profile derived in \cite{Green17}, or more precisely, its numerical analogous for a 3D ${\bf Q}$ tensor (see \ref{sec:3DNumSolution}), and the numerical results reported in \cite{Marenduzzo07}. 
We quantify the deviations from the theoretical prediction, Eq. (\ref{Vel_analytical}), through the ratio between the maximum magnitude of the velocity and the maximum of the analytical profile: when the solution deviates from (\ref{Vel_analytical}) this quantity departs from unity.

In Fig. \ref{2D_map} (a) and (b) we show the behavior of the rescaled maximum magnitude of the velocity in logarithmic scale for negative and positive values of the activity parameter as a function of $\Pi_3$ and $|\Pi_1|$ with $\Pi_3$ ranging over almost four order of magnitudes: $6.67<\Pi_3<2 \cdot 10^4$, and $|\Pi_1|$ spanning over six order of magnitudes: $20<|\Pi_1|<2 \cdot 10^7$. As a comparison, in \cite{Marenduzzo07} $125<|\Pi_1|<750$ and $\Pi_3\approx0.45$, while in \cite{Shendruk17} $50<\Pi_1<800$ and $\Pi_3\approx0.23$, hence in these studies $|\Pi_1|$ spans at most one order of magnitude within a range we are also covering while $\Pi_3$ is fixed, smaller than the values we select and its effect is not assessed. We explore such a wide range of parameter space to capture both the small and large activity range and include both the FDL regime and a range of parameters where the velocity field has the ability of distorting the nematic profile. The lower boundary for the $\Pi_3=\Gamma\eta$ range is limited by the computational cost of simulations. We have run simulations at least up to a time $T_{final} \approx \tau=L^2/\Gamma K$, sufficient to ensure convergence to a steady state if it exists. We have observed that there is no possibility to reach a steady state for an extensile nematics, $\alpha>0$, at large $\Pi_1$ and away from the FDL ($\Pi_3\ll1$), here solutions remain unsteady as marked in Fig. \ref{2D_map} (b). We stress that even in the flow-tumbling regime we obtain steady state profiles as reported in \cite{Marenduzzo07} rather than oscillatory solutions, as for example in \cite{Thampi15}.

In Fig. \ref{2D_map} (a) and (b) the large $\Pi_3$ and low $|\Pi_1|$ region where the solution is given to a very good approximation by equation (\ref{Vel_analytical}) is conveniently identified by a vanishing small magnitude, outside this area the numerical solution deviates from (\ref{Vel_analytical}) differently for negative and positive $\alpha$. In particular, for large negative values of the $\Pi_1$ parameter the flow is suppressed, while, for large positive values the behavior becomes unsteady. For positive intermediate values of $\Pi_1$ large velocities develop as signaled in the right panel by a dark blue band that bends toward larger $\Pi_3$ for larger $|\Pi_1|$, qualitatively these solutions correspond to those reported in Fig. 14 and 15 in \cite{Marenduzzo07}. 

Fig. \ref{2D_map} (c) and (d) show how the velocity profiles change with $\Pi_3$ for a fixed negative and positive value of the $\Pi_1$ parameter. For negative $\alpha$ the rescaled velocity magnitude decreases with $\Pi_3$ while, in parallel, the velocity profile becomes more and more asymmetric: the positive peak moves toward the wall with parallel anchoring while the negative peak flattens; the trend continues until for the smallest $\Pi_3$ the velocity vanishes. For positive $\alpha$ the profile changes as $\Pi_3$ decreases, from the analytical result, Eq. (\ref{Vel_analytical}), to an either entirely positive or negative one of larger magnitude (the sign is randomly selected by the system), in this configuration the peak is roughly located in the middle of the domain. As $\Pi_3$ is further decreased the rescaled velocity magnitude is reduced, sharper and multiple peaks appear until the profile becomes unsteady. To provide an  overall view on the structure of the active nematics, panels (e)-(g) in Fig. \ref{2D_map} represent the director field in the channel for 3 calculations of map (a)-(b) as indicated by the plot titles. Case (e) corresponds to the analytical solution, Eq. (\ref{analytical_nx_ny}).  

We have verified the sensitivity of the steady state solutions to different initial conditions by repeating the calculations of Fig. \ref{2D_map} with different initial ${\bf n}$-profiles as detailed in the caption of Fig. \ref{bistability}. We find some dependence on the initialization for the $\alpha>0$ solutions with values of the parameters that lie in the parameter-space region located in between the FDL and the unsteady solutions, see Fig. \ref{bistability} for two representative examples. As expected, no dependence on the initial conditions is found in the FDL region, as well as in the $\alpha<0$ semi-plane of parameter space.

The results reported in Fig. \ref{2D_map} are obtained for a flow aligning nematic, $\xi = 0.7$. For the flow tumbling regime, e.g. $\xi=0.3$, the results differ:  the velocity displays a behavior similar to panel (a) of Fig. \ref{2D_map} for both positive and negative values of the activity parameter and the rescaled velocity profiles vary with $\Pi_3$ similarly to panel (c) of Fig. \ref{2D_map}, see Fig. \ref{2D_map_tumbling} (a)-(d). More precisely, although even for the flow tumbling case there are quantitative differences between the results for a negative and positive activity parameter evident by comparing panel (a) and (b) of Fig. \ref{2D_map_tumbling}, qualitatively, an increment in the magnitude of activity or a decrease in $\Pi_3$ leads to a suppression of the flow field. Similarly to the flow aligning case, we also note some instabilities of the numerical solution for large positive values of the activity parameter in the bottom right corner of Fig. \ref{2D_map_tumbling} (b). An interpretation of the differences between the flow aligning and flow tumbling case is provided in the following section, Sec. \ref{IntRes}. 

For both the flow aligning and flow tumbling case the effect of decreasing the $\Pi_3$ parameter is similar to that of increasing the $|\Pi_1| $ parameter, hence for both positive and negative values of activity the smooth transition from the frozen director limit regime occurs along lines of constant $ |\Pi_1|/\Pi_3=|\alpha| L^2/(\Gamma \eta K)$. We draw two of them in Fig. \ref{2D_map} (a)-(b) and Fig. \ref{2D_map_tumbling} (a)-(b): one for $|\Pi_1|/\Pi_3=20$ marking the deviation from solution (\ref{Vel_analytical}) and one for $|\Pi_1|/\Pi_3=2 \cdot 10^{4}$ signaling a second transition to the zero velocity or the unsteady behavior for $\alpha>0$. Given the relevance of the $ \Pi_1/\Pi_3$ dimensionless group we will from now on refer to it with the new symbol $\Phi:=\Pi_1/\Pi_3$. Note that in Fig. \ref{2D_map_tumbling}(c)-(d) we report the velocity profiles for calculations that in the (a)-(b) maps lie along a line of maximum variation of $ |\Phi|$, that is a line perpendicular to the $ |\Phi|$ isolines, rather than on a vertical cut as in Fig. \ref{2D_map}(c)-(d). We have verified in Fig. \ref{collapsed_prof} that for the same value of $ |\Phi|$ we obtain the same director field profile ${\bf n}(y)$ and the velocity profiles collapse on a single curve provided that they are rescaled by the activity parameter $\alpha$. In conclusion, we have hypothesized in Sec. \ref{DimN} that solutions would depend on three parameters: $\Pi_1$, $\Pi_3$ and $\xi$, and we have found numerically that results practically depend on two parameters $\Phi$ and $\xi$.

For the flow tumbling case the profiles for the positive/negative $ \Phi$ appear flipped left to right, top to bottom, we will provide an explanation for this in section \ref{IntRes}. Note that since we have collapsed two dimensionless parameters into one this allows to reproduce the solutions in Fig. 14 and 15 of \cite{Marenduzzo07} using larger $\Pi_3$ values if this is compensated by smaller $|\Pi_1|$ and if the remaining dimensionless numbers, specifically $\Pi_2$ are the same. This reduction of parameter space also explains why in Sec. \ref{sec:3DNumSolution} solution (\ref{Vel_analytical}) was found to a very good approximation up to $|\Pi_1|=10^3$, in that specific case in fact $\Pi_3\approx 10^4$ making the $|\Pi_1|=10^3$ threshold equivalent to $\Phi\approx 0.1<1$.

As a final observation we note a qualitative similarity between the velocity profiles for a contractile nematics for both the flow tumbling and aligning regime, compare Fig. \ref{2D_map}(c) and Fig. \ref{2D_map_tumbling}(c), this suggests that those solutions may have only a weak dependence of the flow aligning parameter $\xi$.

\subsection{Interpretation of results}\label{IntRes}

\begin{figure} \centering
\resizebox{8cm}{!}{\includegraphics{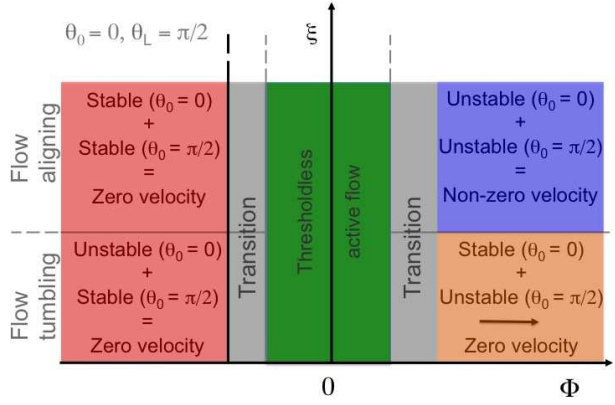}}
\caption{\label{scheme_interpretation}
Schematic representation of the numerical results reported in section \ref{2D_map_results}. For small values of the dimensionless parameter $\Phi = \Pi_1/\Pi_3$ we observe thresholdless active flows \cite{Green17} independently of the sign of the activity parameter or the value of the flow-aligning parameter (green shaded area). For intermediate values of $|\Phi|$ the velocity field is non-zero and depends on the sign of the activity parameter as well as the value of the flow aligning-parameter (grey shaded area). There exists a very close resemblance of the velocity profiles in the transition region for three cases over four: positive activity and flow tumbling Fig. \ref{2D_map_tumbling}(d), negative activity and flow tumbling Fig. \ref{2D_map_tumbling}(c), negative activity and flow aligning Fig. \ref{2D_map}(c). The positive activity and flow aligning case differs and shows some dependence on the initial conditions, Fig. \ref{2D_map}(d). The behavior  in the regions of large magnitude of $\Phi$, beyond the transition regions, can be rationalized on the basis of previous studies \cite{Voituriez05, Ramaswamy07, Edwards08} for  the flows of active nematics in channels with either homeotropic  or paralallel boundary conditions. The hybrid boundary condition ($\theta_0=0$, $\theta_L=\pi/2$) can be interpreted as a combination of parallel and perpendicular boundary conditions, for those cases the stability conditions have been derived in the literature and their combination suggests the type of flow we observe in the hybrid case (blue and red shaded areas).}
\end{figure}

\begin{figure} \centering
\resizebox{18cm}{!}{\includegraphics{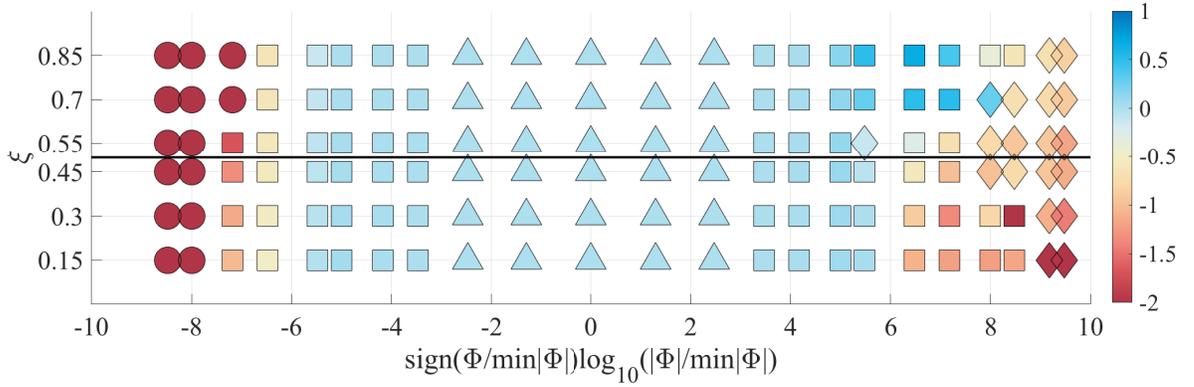}}
\caption{\label{phase_diag_num} Map of the numerical solutions for the 1D velocity profile as a function of the flow-aligning parameter $\xi$ and the dimensionless number $\Phi$. Different symbols correspond to different type of solutions: circles denote zero-velocity solutions that correspond to a diagonal ${\bf Q}$-tensor; triangles indicate solutions whose normalized root mean square (RMS) value does not deviate from Eq. (\ref{Vel_analytical}) by more than 0.01; squares indicate all the other type of solutions and diamonds designate unsteady solutions. The color represents the rescaled magnitude of velocity as in Fig. \ref{2D_map}(a)-(b) and \ref{2D_map_tumbling}(a)-(b). The boundary between the flow-tumbling and flow-aligning behavior is given by $\xi=0.5$ and is marked by a black line. Note the expression for the $x-$axis chosen to display $\Phi$ in $\log$-scale while distinguishing between the positive and negative cases.}
\end{figure}

A stability analysis performed on the $\bf{n}$-model \cite{Voituriez05} and later results \cite{Edwards08} built on expanding concepts presented in \cite{Ramaswamy07} show that for a 1D slab geometry in a flow aligning regime a nematic profile parallel to the walls is ({\it i}) unstable for extensile active particles ($\alpha>0$)  and ({\it ii}) stable for contractile ones ($\alpha<0$). Similarly, a nematic arrangement perpendicular to the walls is ({\it iii}) unstable for extensile active particles, ({\it iv}) stable for contractile ones. In these cases the instability appears above a certain activity threshold $\alpha_c$ that depends on several model parameters: the system size $L$, the dynamic viscosity of the flow $\eta$, the elastic, flow aligning and rotational diffusivity parameter, see {\it e.g.} \cite{Edwards08} for an analytical expression for $\alpha_c$. We have verified numerically that this critical threshold also predicts the transition to spontaneous active flows in the ${\bf Q}$-model when the anchoring is forced through Dirichlet boundary conditions $n_x = \cos\theta$, $n_y = \sin\theta$. In \cite{Edwards08} free boundary conditions were imposed for the director field at the walls ($\partial_y n_i = 0$).

Hybrid boundary conditions can be viewed as a combination of the four scenarios ({\it i}-{\it iv}), both a nematic arrangement parallel and perpendicular to the walls are unstable and will result in a non-zero flow for positive and large enough $\alpha$ while instabilities are suppressed for both these configurations when $\alpha<0$. This explains why a deviation from the FDL will lead for $\alpha<0$ to a suppression of the flow and for $\alpha>0$ to a non-zero velocity profile providing an explanation for the behavior reported in Fig. \ref{2D_map} (a) and (b). More in detail, the analogy with the homogeneous director field can be thought to  hold {\it locally} according to the mechanism described in \cite{Ramaswamy07}. In contractile systems both the parallel and perpendicular nematic arrangements are stable. Therefore if in the neighbourhood of the walls where the anchoring is fixed the nematic profile is distorted from the parallel or perpendicular alignment such distortions will die away. Two separate domains will form, one with uniform $n_x=1,$ one with uniform $n_y=1$, a discontinuity in the ${\bf n}$ profile will appear where $q_0$ will modulate its magnitude and go to zero, see Sec. \ref{FRzfs}. See Fig. \ref{scheme_interpretation} for a schematic representation of our interpretation of the results. 

The situation is different for the flow tumbling regime: in this case a nematic arrangement parallel to the walls is stable for extensile active particles and unstable for contractile ones, on the contrary, a nematic arrangement perpendicular to the walls is unstable for extensile active particles and stable for contractile ones \cite{Edwards08}. Hence mixed boundary conditions in the flow tumbling regime correspond to the combination of a stable and unstable configuration in which the stable tendency wins over the unstable one leading to zero-velocity profile as the magnitude of activity increases, see the schematic representation in Fig. \ref{scheme_interpretation}. 

For intermediate values of the activity parameter in the flow-tumbling regime the velocity profiles for $\alpha>0$ closely resemble the velocity profiles for the $\alpha<0$ case once `flipped' about the $y$-axis, this symmetry reflects the symmetries embodied in the equations for the nematic field as stressed in \cite{Edwards08}: a change in the sign of $\alpha$ is equivalent to a change in sign of the flow-aligning parameter in conjunction with a $\pi/2$ rotation of the director field. Therefore changing the sign of $\alpha$ in our setting is equivalent to exchanging the $y=0$ and $y=L$ boundary conditions as emerges also from Fig. \ref{collapsed_prof} (c)-(d).

Figure \ref{phase_diag_num} displays on the $\xi$-$\Phi$ plane the different type of solutions described in this work for the same $\Phi$ values of Fig. \ref{2D_map}, \ref{2D_map_tumbling} and some additional $\xi$ values. This plot corresponds to the numerical outcome and corroborates the schematic representation of Fig. \ref{scheme_interpretation}.

\subsubsection{Further remarks on the zero-flow solution}\label{FRzfs}

The nematic profile selected dynamically by the system and associated to the zero-flow steady state is a free energy stationary point that satisfies $\mathcal{H}_{ij}=0$ and corresponds to a zero-curl active force. This second condition is verified in our setting anytime the off-diagonal terms of the $Q_{ij}$ tensor are zero. The nematic profile will therefore satisfy an undamped unforced Duffing equation: $KQ''_{xx}-AQ_{xx}-2CQ_{xx}^3=0$ in 2D and the system of non-linear ODEs 
$$Q''_{xx} = \frac{1}{K}[AQ_{xx}+BQ_{xx}^2+2C(Q_{xx}^2+Q_{yy}^2+Q_{xx}Q_{yy})Q_{xx}-2B(Q_{xx}^2+Q_{yy}^2+Q_{xx}Q_{yy})/3],$$
$$Q''_{yy} = \frac{1}{K}[AQ_{yy}+BQ_{yy}^2+2C(Q_{xx}^2+Q_{yy}^2+Q_{xx}Q_{yy})Q_{yy}-2B(Q_{xx}^2+Q_{yy}^2+Q_{xx}Q_{yy})/3]$$ in 3D.
The solutions obtained with hybrid anchoring boundary conditions for $\Pi_2\ll1$ are characterized by sharp fronts in the $q_0$ profile where $q_0\rightarrow0$ while ${\bf n}$ changes orientation to match the boundary conditions switching from $n_x=1$, $n_y=0$ to $n_x=0$, $n_y=1$  [see Fig. \ref{2D_map_tumbling}(e) blue curve]. In the 3D case biaxiality develops in the region where ${\bf n}$ changes orientation (see Sec. \ref{BI}). These solutions reflect the greater generality of the ${\bf Q}$-model, in fact they are not admitted in the ${\bf n}$-model where the magnitude of the nematic order parameter is fixed. In conclusion, we discover that in addition to Eq. (\ref{analytical_nx_ny}), that we refer to as fixed point 1 (FP1), the Euler-Lagrange equation for the $\bf{Q}$-model admits a second stable fixed point, FP2, that allows for steady state zero-flow solutions that manifest at non-zero activity. In the following section we deepen our analysis on these two configurations.
 
 \subsection{Comments on the Minimum Energy solutions in the Q model\label{GS}} 
 
\begin{figure} \centering
\resizebox{15cm}{!}{\includegraphics{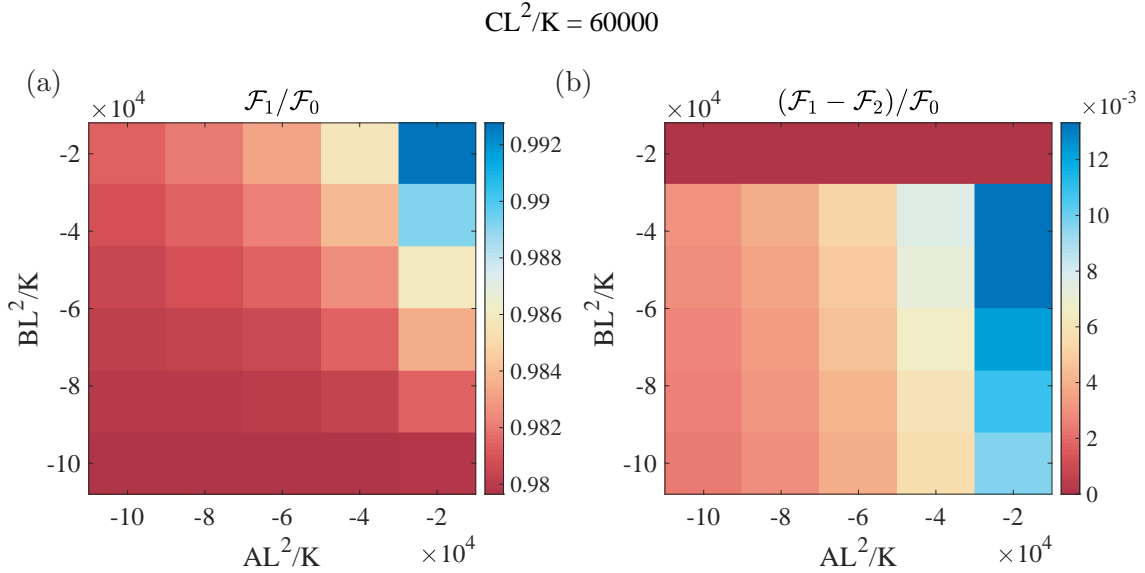}}
\put(-420,180){(a)}
\put(-220,180){(b)}
\caption{\label{GS_diagrams} (a) Rescaled Landau-de Gennes free energy as a function of the dimensionless parameters $BL^2/K$ and $CL^2/K$ for the solution of $\mathcal H_{ij} = 0$ obtained by relaxing the initial condition IC1 through Eq. (\ref{relaxation}). The free energy is rescaled by a reference free energy value, $\mathcal{F}_{0}$, corresponding to an homogeneous solution for a system of size $L$. (b) Difference between the Landau-de Gennes free energy associated to IC1, $\mathcal{F}_{1}$, and IC2, $\mathcal{F}_{2}$, rescaled by $\mathcal{F}_{0}$. These calculations have been performed for $CL^2/K=60000$ and repeated for $0<CL^2/K<100000$. In this parameter range we observe qualitatively the same type of solutions, differences concern the magnitude of $q_0$ in a narrow region close to the boundaries.}
\end{figure}

 In Sec. \ref{sec:3DNumSolution} the numerical solution for a 3D-${\bf Q}$ tensor, FP1, was obtained for values of the thermotropic constants and $q_0$ that corresponded to minimum energy solutions for uniform states ($q_0=constant$), see caption of Fig. \ref{rms_dev}. These same values were used when integrating the full set of equations leading to the dynamical selection of FP2 for low $\Phi$. We now test the sensitivity of the two fixed points to the parameters $A$, $B$, and $C$, by looking for solutions of $\mathcal H_{ij} = 0$ in a neighbourhood of the previously selected values: 
 we vary $A$ and $B$ in the range $-1.67 \le A/C \le1.67$ and $-1.67 \le B/C \le 0.0$ with $CL^2/K=6\cdot10^4$.   We always constrain the choice of parameters to thermodynamically stable states ($C>0$) \cite{IntroSM}. Numerically, we find minimum energy solutions relaxing  the order parameter through Eq. (\ref{relaxation}) with fixed anchoring at the walls. We repeat the calculations for two different initial conditions: expression (\ref{analytical_nx_ny}), referred to as `IC1', and a discontinuous initial state with $n_x(y)=1$ for $y=[0, L/2)$, $n_y(y)=1$ for $y=[L/2, L]$, IC2. As expected \cite{Mottram14}, the solution converges to a nematic state for $A<0$ and an isotropic state for $A>0$. The isotropic state is only attained in the middle of the domain given the fixed anchoring at the walls. For $B/C<0$ and IC1 the nematic solution corresponds to a nematic state with a non-zero curl active force of the kind reported in Sec. \ref{sec:3DNumSolution}: the nematic profile corresponds to Eq. (\ref{analytical_nx_ny}) to a very good approximation while the $q_0$ profile slightly changes as a function of the thermotropic parameters. In Fig. \ref{GS_diagrams} (a) we show the free energy of this solution as a function of the thermotropic parameters and we find that when compared to a uniform state solution the most energetically favorable configurations are attained for the largest $AL^2/K$ and $BL^2/K$. For $B/C<0$ and IC2 the solution is a nematic state with a zero-curl active force and corresponds to FP2.  For the special case $B=0$ both IC1 and IC2 converge to FP2.  When we compare the Free Energy value for the solutions obtained with IC1 and IC2, Fig. \ref{GS_diagrams} (b), we find that FP2 has the largest energy hence is a local minimum.
 
 \subsection{An example of biaxial thresholdless active flows in the ${\bf Q}$ model\label{BI}}
 
\begin{figure} \centering
\resizebox{7cm}{!}{\includegraphics{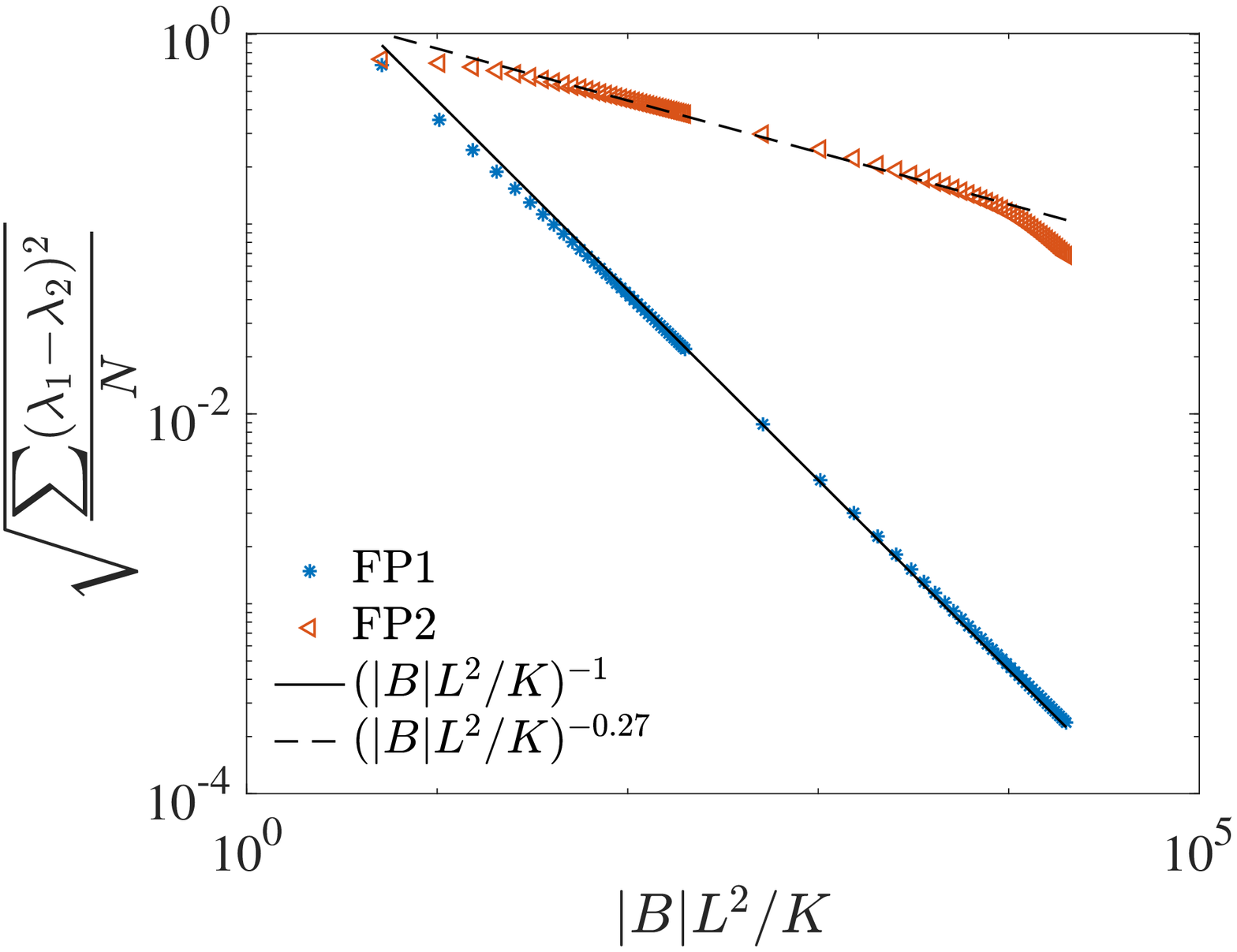}}
\resizebox{8cm}{!}{\includegraphics{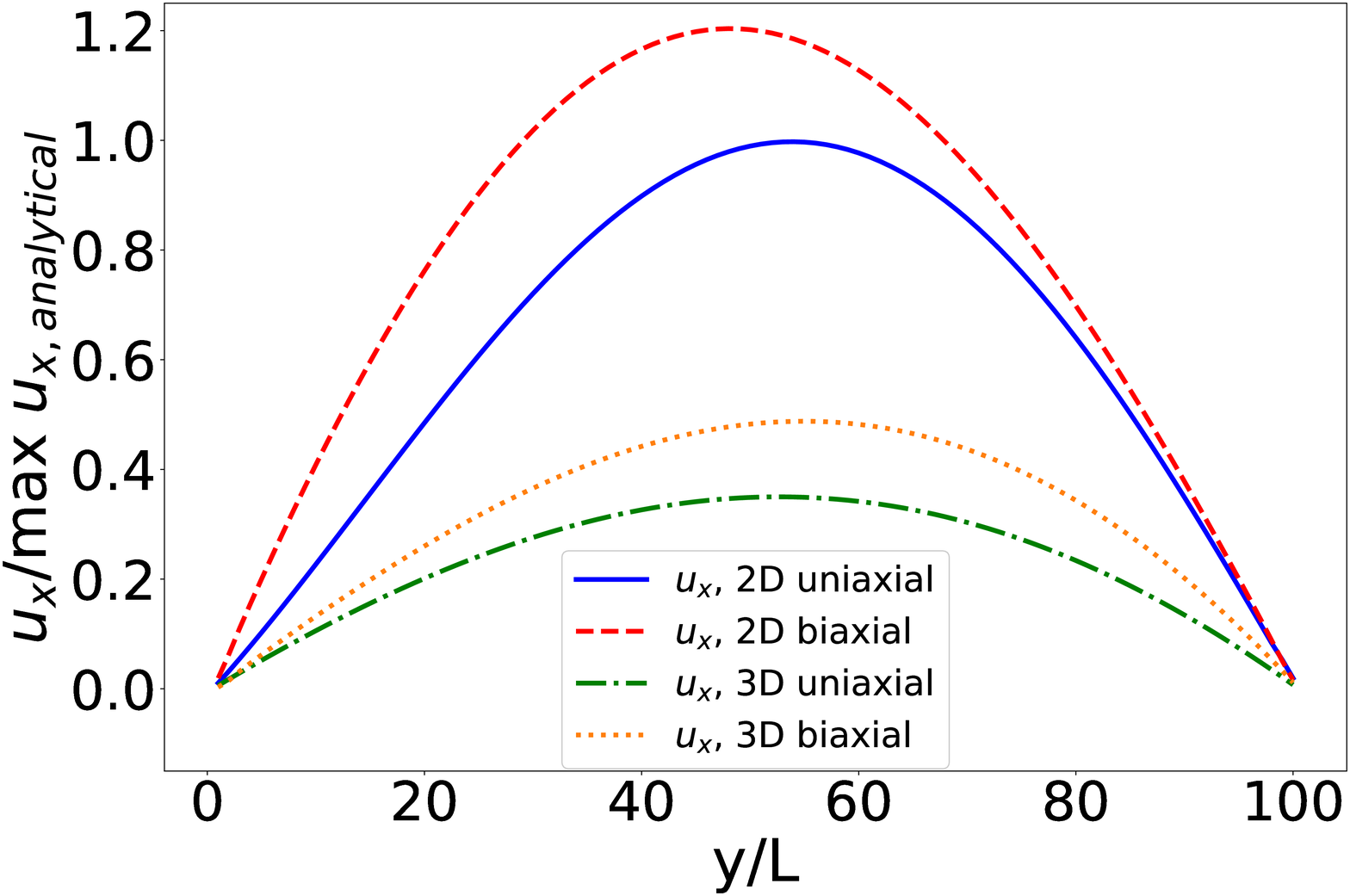}}
\put(-450,150){(a)}
\put(-220,150){(b)}\\
\resizebox{8cm}{!}{\includegraphics{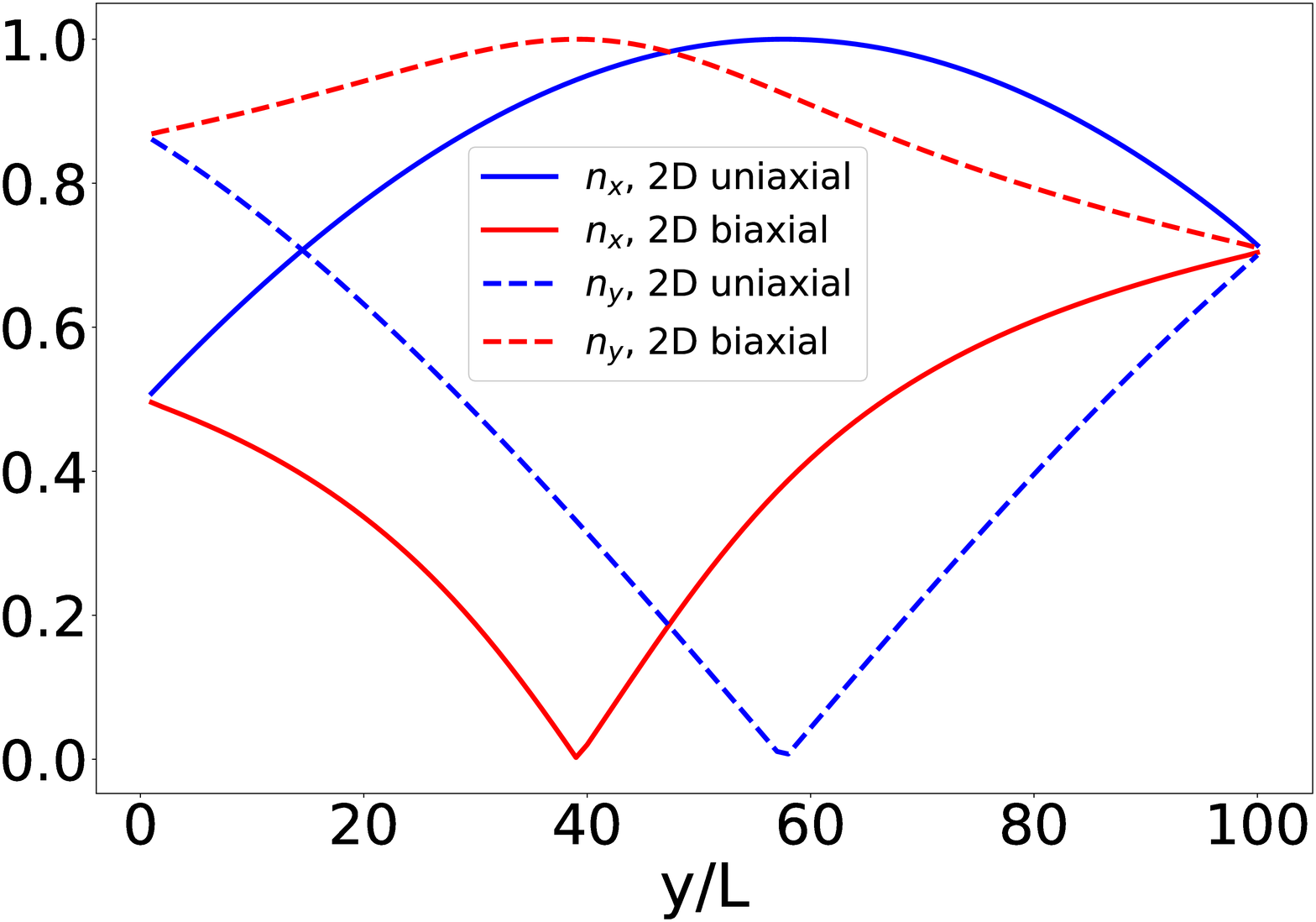}}
\resizebox{8cm}{!}{\includegraphics{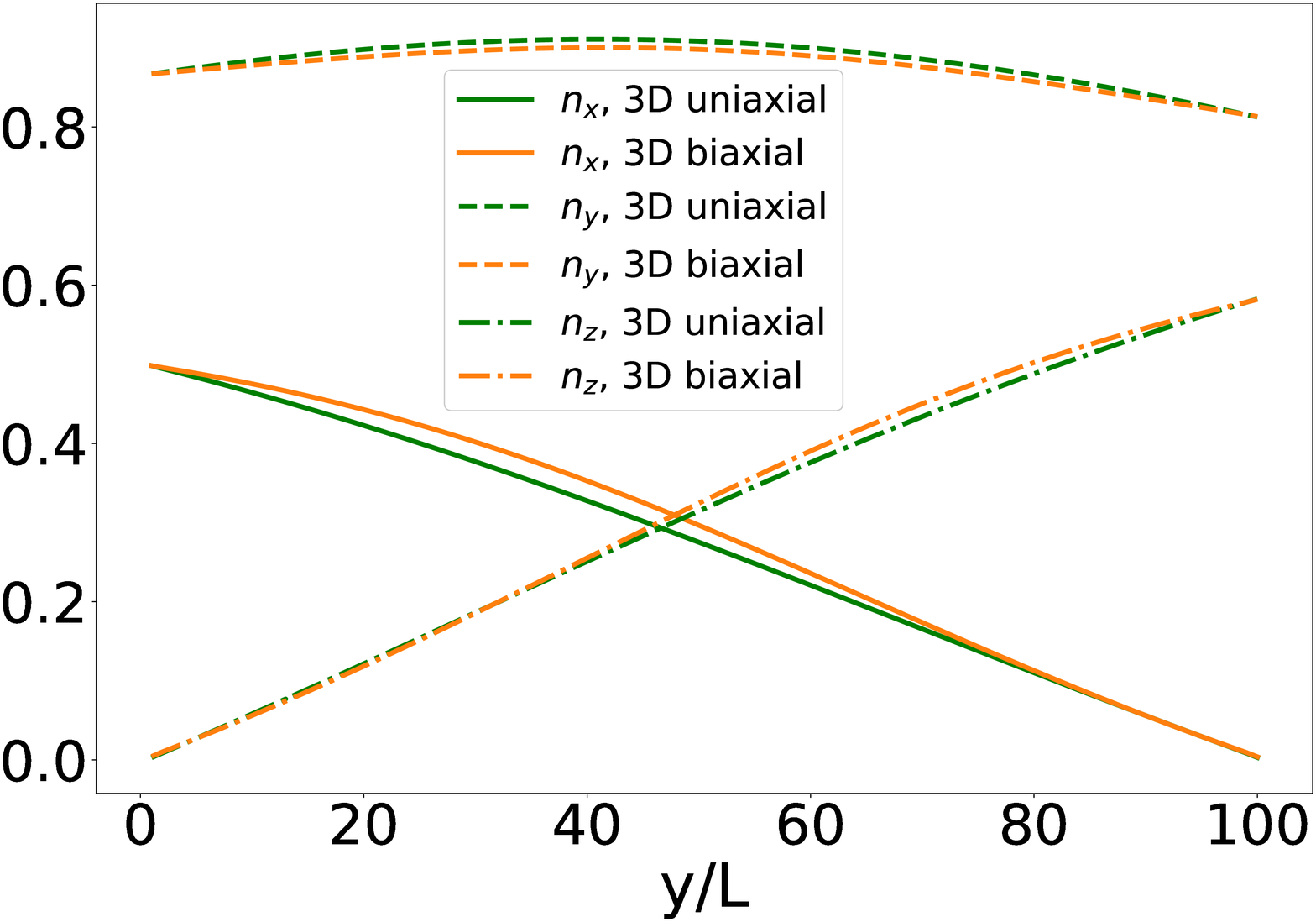}}
\put(-450,140){(c)}
\put(-220,140){(d)}
\caption{\label{Biaxial_sol} (a) root mean square deviation from uniaxiality measured in terms of the difference between the closest eigenvalues of the $Q_{ij}$ tensor for FP1, FP2 and $BL^2/K\rightarrow 0$, here $CL^2/K = |A|L^2/K =  60000$. The solid and dashed lines represent power-law scalings as reported in the legend. (b) velocity profile for a thresholdless active flow for a 2D and 3D geometry and for a uniaxial case (ground state of type 3) and biaxial case (ground state of type 4). The 2D uniaxial profile is given by Eq. (\ref{GenVitSol}), the 3D velocity field has a non-zero $u_z$ component that is not reported in the plot for clarity. All the curves are rescaled by the maximum of the analytical profile (\ref{GenVitSol}). In panel (c) and (d) we report the nematic director field for the 2D (c) and 3D (d) solution.}
\end{figure}

The Landau-de Gennes free energy adopted in the ${\bf Q}$-model [Eq. (\ref{FLdG})] allows for a wider family of minimum energy solutions than the Frank free energy used in the ${\bf n}$-model because it includes a thermotropic term in addition to a distortion term. This fact is relevant when dealing with thresholdless active flows since they require minimum energy nematic profiles with a non-zero curl active force. Potentially, the ${\bf Q}$-model allows for more thresholdless active flow configurations than the ${\bf n}$-model including, in 3D, biaxial solutions. The objective of this section is to identify some of them.
 
To decide whether the degree of biaxiality of a solution is non-negligible a threshold is set on the difference between the two closest eigenvalues $\lambda_i$, $\lambda_j$ of tensor ${\bf Q}$: if $\sqrt{\sum_k (\lambda_i-\lambda_j)^2}/L>10^{-4} $, where $L$ is the system size, and $k$ are the grid points, then biaxiality is considered non-negligible. 
 
 We classify the type of minimum energy solutions that we obtain in five categories: isotropic states (type 0), nematic states with a zero-curl active force $\nabla \times f_a=0$ and negligible biaxiality (type 1), biaxial nematic states with a zero-curl active force (type 2), nematic states with a non-zero curl active force $\nabla \times f_a \ne 0$ and negligible biaxiality (type 3), and biaxial nematic states with a non-zero curl active force (type 4).  Only solutions of type 3 and 4 can support thresholdless active flows. The two fixed point solutions discussed so far correspond to type 3, FP1, and type 2, FP2. A closer inspection of FP1 reveals that this fixed point solution has non negligible biaxiality in the nieghborhood of $BL^2/K=0$. Similarly, the biaxial fixed point FP2 more markedly deviates from a uniaxial arrangement as $BL^2/K\rightarrow0$, see Fig. \ref{Biaxial_sol}(a). The RMS deviation from biaxiality for both FP1 and FP2 follows a power law as reported in Fig. \ref{Biaxial_sol}(a). 
 
The only biaxial solution identified so far for $BL^2/K=0$, FP2, will not be able to sustain a thresholdless active flow, however, different anchoring choices will change this picture. For example, for anchoring angles of 60 and 45 degrees and $B = 0$, the ground state is biaxial and has a non-zero curl active force, therefore supports a biaxial thresholdless active flow, Fig. \ref{Biaxial_sol} (b)-(d). If this geometry is extended in 3D meaning that the plane formed by the anchoring angles at the walls is not orthogonal to the walls (the angle $\theta(0)$ lies on the $x$-$y$ plane, the angle $\theta(L)$ lies on the $y$-$z$ plane) the picture is similar, Fig. \ref{Biaxial_sol} (a)-(d). Note that for $B = 0$ the isotropic-nematic phase transition is second order instead of first order and consider that the degree of biaxiality grows as the value of the $B$ parameter approaches zero Fig. \ref{Biaxial_sol}. We can therefore conclude that biaxiality is relevant for a weakly first order isotropic-nematic phase transition.
 
\section{Conclusions}\label{Conclusions}
We study active nematic flows confined in a quasi one-dimensional channel geometry with hybrid alignment at the walls, more specifically, we impose a fixed anchoring parallel to one wall and perpendicular to the second. Active flows in this setting have been investigated in previous studies revealing interesting features: in \cite{Marenduzzo07} it was shown how small positive and negative values of the activity parameter lead to different velocity profiles while in \cite{Green17} it was demonstrated that this geometry present a non-zero velocity field even for vanishingly small values of the activity parameter. In \cite{Green17} an analytical solution for such a {\it thresholdless} active flow was derived within the active nematohydrodynamic ${\bf n}$-model for small activity and in the frozen director limit (FDL), that is in a regime where the nematic is not distorted by the flow and satisfies the Euler-Lagrange equation for minimizing the free energy. In this paper we have shown that this solution holds also in the active nematohydrodynamic ${\bf Q}$-model for a two-dimensional ${\bf Q}$-tensor, a result that can be generalized to any anchoring angle. We reproduce this solution numerically with an hybrid Lattice-Boltzmann code identifying the range of model parameters for which this result is found with high accuracy. In addition, we verify numerically that this nematic and velocity profile is a very good approximation of the solution for a three-dimensional ${\bf Q}$-tensor. 

The active nematohydrodynamic ${\bf Q}$-model generally depends on 6 dimensionless numbers. However, in our specific geometry, in the absence of an external forcing, and for system sizes much larger than the characteristic defect core we expect the solution to depend on three dimensionless groups: $\Pi_1$, that is the square of the ratio between the active length scale and the size of the system, $\Pi_3$, a parameter that measures the distance from the FDL regime identified by $\Pi_3\gg1$, and the flow aligning parameter $\xi$ that expresses the tendency of particles to tumble or align with the flow. With the aim of providing a unifying picture for active flows in a channel with hybrid anchoring at the walls, we have computed numerically steady state solutions in a wide portion of parameter space: the parameter $|\Pi_1|$ spans six orders of magnitude, the parameter $\Pi_3$ spans almost four orders of magnitude, while the values of $\xi$ encompass both the flow-tumbling and aligning regime. These parameter ranges include both the FDL and a parameter region where the velocity field has the ability of distorting the nematic profile and comprise both the small and large activity limit.
Our study reveals that the effect of decreasing $\Pi_3$ is similar to that of increasing $|\Pi_1|$ so that the transition from the FDL solution occurs along lines of constant $\Phi=\Pi_1/\Pi_3$, hence, the solution only depends on 2 dimensionless groups: $\Phi$ and $\xi$, a result that could not be anticipated theoretically.

We observe that the symmetric thresholdless active flow derived in \cite{Green17} manifests to a very good approximation for small values of $\Phi$ and is independent of the sign of activity and the value of $\xi$. Moving away from the low $\Phi$ region the transition from the symmetric active flow is smooth with the model parameters and depends on them. In particular, for ({\it i}) a flow-aligning and contractile nematic the velocity profile becomes more and more asymmetric while its magnitude rescaled by the activity parameter decreases as $|\Phi|$ increases until the flow is completely suppressed, for ({\it ii}) a flow-aligning and extensile nematic the velocity profile loses symmetry as $\Phi$ increases until it becomes unsteady, for intermediate values of the parameter the velocity profile has a single peak located around the middle of the domain, this steady state configuration displays some dependence on the initial condition for the nematic director field. Unlike in the flow-aligning regime, in the flow-tumbling regime the deviation from the FDL profile is similar for a ({\it iii}) contractile and ({\it iv}) extensile nematic, in both cases the profile decreases in relative magnitude until the flow is suppressed as $|\Phi|$ increases.

We interpret the different flow-aligning and tumbling behaviors for large $|\Phi|$ in terms of the stability of simpler configurations with either parallel or perpendicular anchoring at both walls \cite{Edwards08}. In the flow-aligning regime both a parallel and perpendicular configuration is stable to perturbations for negative activity and unstable for positive activity, this provides a rationale for the zero-flow solution observed for a contractile nematic with hybrid anchoring at the walls and the non-zero large magnitude or unsteady velocity solution found for an extensile nematic with hybrid anchoring. In the flow tumbling regime the picture is different, for a contractile particle the flow is unstable to perturbation for a parallel arrangement and stable for a perpendicular one while the opposite is true for extensile active particles. This means that mixed boundary conditions correspond to a combination of a stable and unstable configuration for both negative and positive activity and we observe that the stable tendency wins over the unstable one leading to zero-flow solutions qualitatively very similar to the zero-flow solutions found for the contractile flow-aligning case. Therefore, unlike in the flow-aligning case, in the flow-tumbling case there is a symmetry in the behavior for positive and negative value of activity.

In the zero-flow configuration the nematic director ${\bf n}$ reorients abruptly from $n_x=1$, $n_y=0$ to $n_x=0$, $n_y=1$ to match the boundary conditions while $q_0$ decreases to zero in correspondence of the discontinuity. We clarify that this configuration supports a zero-flow steady state because it corresponds to a local minimum of the free energy and a zero-curl active force. We have so found a second stationary point for the free energy in addition to the nematic profile responsible for the thresholdless active flow. This stationary point is a local minimum of the Euler-Lagrange equation and displays biaxiality for a three-dimensional $\bf Q$-tensor.

Finally, we exploit the greater generality of the ${\bf Q}$-model compared to the ${\bf n}$-model and provide an example of a biaxial thresholdless active flow for conflicting anchoring at the walls corresponding to a 60 and 45 degree angle on either a two-dimensional or three-dimensional (out of the plane) geometry. For this configuration the biaxial thresholdless flow exists also in the special case of a symmetric quartic free energy expression that corresponds to a second-order isotropic-nematic phase transition. In our examples we find that biaxiality is relevant for a weakly first-order isotropic-nematic phase transition.

As a concluding remark we remind the reader that our results have been obtained in a one-dimensional domain as representative of two-dimensional channel flows that are uniform along the longitudinal direction. We recall that in a truly two-dimensional system instabilities can develop in the longitudinal direction due to spontaneous symmetry breaking, see for example \cite{Shendruk17, Chandragiri19}. Therefore, an important underlying question is the range of validity of our analysis when extended to 2D systems. Informed by the results of our study, we expect the critical longitudinal wave length to depend on two parameters: $\Phi$ and $\xi$. Preliminary results point to the fact that lower is the value of the parameter $\Phi$, more robust is the 1D approximation, or else, longer is the critical longitudinal wave length, $\lambda_{x, c}$. Assessing the role of the flow-aligning parameter $\xi$ proves to be more difficult. Addressing the functional form of $\lambda_{x, c}$ is by itself a relevant and complex matter that will be the subject of future studies.

\section*{Acknowledgements}
{\it This project has received funding from the European Union's Horizon 2020 research and innovation programme under the Marie Sklodowska-Curie grant agreement N 754462.} I.P. acknowledges support from Ministerio de Ciencia, Innovaci\'on y Universidades (Grant No. PGC2018-098373-B-100/FEDER-EU), DURSI (Grant No. 2017 SGR 884), and SNSF (Project No. 200021-175719). C.R. thanks Dr. G. Di Staso and Dr. D. Banerjee for useful scientific discussions and the initial support with the Lattice Boltzmann code.

\newpage\null\thispagestyle{empty}\newpage
\bibliography{ms1.bbl}
\end{document}